\documentclass[aps,prb,nobalancelastpage,longbibliography,showpacs,twocolumn,superscriptaddress,floatfix]{revtex4-1}

\usepackage{hyperref}
\usepackage{graphicx}
\usepackage{amssymb}
\usepackage{amsmath}
\usepackage{eucal}
\usepackage{color}
\usepackage{bm}

\newcommand{\e}{{\rm e}}

\newcommand{\eps}{\epsilon}

\newcommand{\kB}{ k_{\rm B} }

\definecolor{DarkGreen}{rgb}{0,0.7,0}

\begin{document}

\title{
Potential barriers are nearly-ideal quantum thermoelectrics at finite power output}

\author{Chaimae Chrirou}
\affiliation{Universit\'e Grenoble Alpes, CNRS, LPMMC, 38000 Grenoble, France.
}
\affiliation{Laboratory of R\&D in Engineering Sciences, Faculty of Sciences and Techniques Al-Hoceima,\\ Abdelmalek Essaadi University, Tetouan, Morocco
}
\author{Abderrahim El Allati}
\affiliation{Laboratory of R\&D in Engineering Sciences, Faculty of Sciences and Techniques Al-Hoceima,\\ Abdelmalek Essaadi University, Tetouan, Morocco.
}
\author{Robert S.~Whitney}
\affiliation{Universit\'e Grenoble Alpes, CNRS, LPMMC, 38000 Grenoble, France.
}

\date{April 7, 2025}
\begin{abstract}
Quantum thermodynamics defines the ideal quantum thermoelectric, with maximum possible efficiency at finite power output. However, such an ideal thermoelectric is challenging to implement experimentally. 
Instead, here we consider two types of thermoelectrics regularly implemented in experiments: (i) finite-height potential barriers or quantum point contacts, and (ii) double-barrier structures or single-level quantum dots.
We model them with Landauer scattering theory as (i) step transmissions and
(ii) Lorentzian transmissions, respectively.  We optimize their thermodynamic efficiency for any given power output, when they are used as thermoelectric heat engines or refrigerators. 
The Lorentzian's efficiency is excellent at vanishing power, but we find that it is poor at the finite powers of practical interest. 
In contrast, the step transmission is remarkably close to ideal efficiency (typically within 15\%) at all power outputs.  The step transmission is also close to ideal in the presence of phonons and other heat leaks, for which the Lorentzian performs very poorly.  
Thus, a simple nanoscale thermoelectric --- made with a potential barrier or quantum point contact --- is almost as efficient as an ideal thermoelectric.
\end{abstract}

\maketitle

\begin{figure}[b]
\includegraphics[width=\columnwidth]{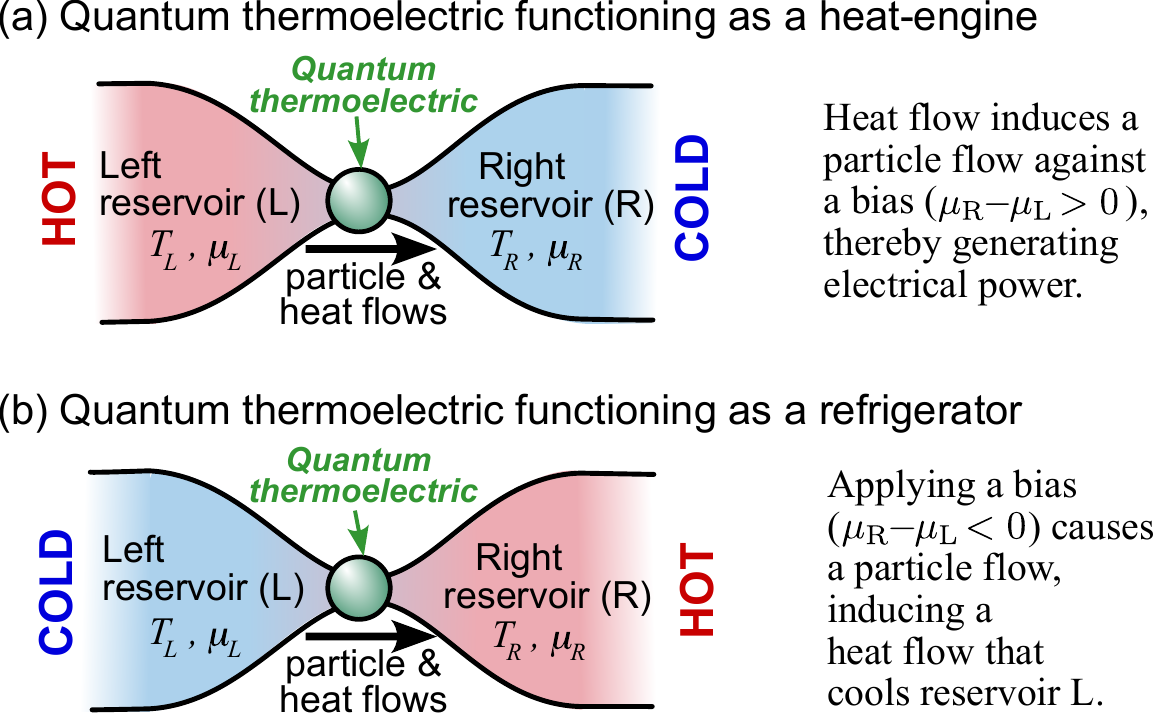} 
\caption{
\label{fig:schematic}
Quantum thermoelectrics couple heat and particle flows between reservoirs. In (a), the heat flow from hot to cold induces a particle flow against a bias (analogous to pushing particles up a hill) so it generates electrical power. In (b), the bias creates a particle flow 
(particles flowing down hill), this induces a heat flow from cold to hot, which cools the cold reservoir. This work takes thermoelectrics that have been demonstrated experimentally, and optimizes their efficiency for any given power output.
}
\end{figure}

\section{Introduction}

No heat engine or refrigerator can exceed Carnot efficiency \cite{Callen1985Sep}, but the route to achieving Carnot efficiency has been identified for both thermoelectric materials\cite{Mahan1996Jul} and quantum thermoelectrics.\cite{Humphrey2005Mar} However, quantum thermodynamics indicates that Carnot efficiency in thermoelectrics is only attainable at vanishing power output, with a stricter upper limit on efficiency for any finite power output. \cite{Whitney2014Apr,Whitney2015Mar} This applies to both heat engines and refrigerators, see Fig.~\ref{fig:schematic}, and is a consequence of the quantum wave-nature of electrons. For non-interacting electrons,
modeled within Landauer scattering theory, this stricter upper limit depends only on the power output, the reservoir temperatures,
and two fundamental constants ($\kB$ and $\hbar$)\cite{Whitney2014Apr,Whitney2015Mar,Whitney2016May,Ding2023Oct}.
Similar results were found in Boltzmann transport theory\cite{Maassen2021Nov}\footnote{Scattering theory holds for non-interacting electrons, or when inter-particle interactions are treated at a mean-field (Hartree) level. Ref.~[\onlinecite{Maassen2021Nov}] uses the linear Boltzmann transport equation within the relaxation time approximation. The question of a bound on efficiency at finite power remains open in other systems, with this bound exceeded in simulations of some systems with strong relaxation\cite{Brandner2015Jan} or inter-particle interactions.\cite{Luo2018Aug}}
Unfortunately, achieving the upper limit requires the quantum thermoelectric to have a boxcar transmission function --- which ensures that electrons flow only in a specific energy window. While there are theoretical proposals that get fairly close to such a boxcar transmission, \cite{maiti2013mobility,Whitney2015Mar,yamamoto2017thermoelectricity,samuelsson2017optimal,haack2019efficient,chiaracane2020quasiperiodic,haack2021nonlinear,behera2023quantum,brandner2025thermodynamic} 
it is challenging to implement them experimentally.\cite{Chen2018-experiment,reihani2024cooling}

Here we take the opposite approach, by considering simple models of quantum thermoelectrics that are {\it easily implemented in experiments}, to explore how close they could get to this ideal efficiency for any given power output. Our modeling uses Landauer scattering theory, optimized by varying experimentally accessible parameters.  These thermoelectrics fall into two categories. 
\begin{itemize}
\item[(a)] Finite-height potential-barriers that are implemented in experiments
at both low\cite{Mykkanen2020Apr} and ambient temperatures,\cite{Shakouri1997Sep,Shakouri1998Jan} or with quantum point-contacts at low temperatures.\cite{Molenkamp1990Aug,Dzurak1993Oct,Brantut2013Nov,footnote-atoms}
We assume that the potential barrier's height is $\epsilon_0$, and that it is thick enough that no electrons tunnel through it, so all electrons at energies below $\epsilon_0$ are reflected, while all those above $\epsilon_0$ transmit over the barrier.
So it can be modeled as a {\it step-function transmission}. 
\item[(b)] Double-barrier structures implementing single-level quantum dots in experiments at low temperatures\cite{prance2009electronic,Fahlvik2012Mar,Josefsson2018Oct}, or transport through a single level of a molecular junction at ambient temperatures.\cite{Reddy2007Mar,cohen2025unusually}
We model this as a {\it Lorentzian transmission}
centered at the dot's energy level, $\epsilon_0$ with a broadening given that level's coupling to the reservoirs, $\Gamma$. 
\end{itemize}

Historically, many authors considered efficiency without worrying about power output. Then very elegant theoretical works\cite{Mahan1996Jul,Humphrey2005Mar} showed that
the maximum efficiency is Carnot efficiency, achieved by vanishing-width transmission functions\cite{Mahan1996Jul} (delta-function-like), which can be implemented as narrow Lorentzians.\cite{Humphrey2005Mar}  While vanishing width implies vanishingly small power output, there was a general perception that finite-width Lorentzians would be desirable for finite power outputs\cite{edwards1993quantum,Jordan2013Feb}.  
This was reinforced by the proof that boxcar functions give maximum efficiency at finite power\cite{Whitney2014Apr,Whitney2015Mar,Whitney2016May}, since a Lorentzian is similar to a smoothed boxcar function.
This made it natural to suppose that a Lorentzian could be tuned to close to the upper limit on efficiency (given by the ideal boxcar); we show here that this is {\it not} the case.  

Our modeling predicts that the Lorentzian transmission is far from ideal,
except at very small power output.  In contrast, it shows that a step-function transmission
is close to ideal efficiency. For the heat engine, it is 
within 15\% of ideal at all power outputs for a broad range of temperatures (see Fig.~\ref{fig:best-engine}).
For refrigerators, it is mostly also within 15\%  of ideal, although
it is a bit worse at lower cooling powers when the temperature ratio of hot to cold is large
(e.g.~$T_{\rm R}/T_{\rm L} = 4$ in Fig.~\ref{fig:best-fridge}).

The poor performance of the Lorentzian is even more stark in the presence of the heat leaks that occur in any real system due to phonons (or other processes) that carry heat between the hot and cold reservoirs. 
Once such heat leaks are taken into account, the maximum efficiency occurs at finite power output\cite{Whitney2015Mar,Ding2023Oct}, where the Lorentzian performs very poorly, while the potential barrier remains close to ideal.

Our modeling clearly suggest that it would be worthwhile to perform experiments on any system that might have such a step-function transmission, since it could then be close to the ideal thermoelectric.

\section{Power and efficiency in nonlinear scattering theory}

Numerous works on nanoscale thermoelectrics have used scattering theory, many of them are reviewed in sections 4-6 of Ref.~[\onlinecite{Benenti2017Jun}], or Refs.~[\onlinecite{cui2017perspective,bedkihal2025fundamental-review}].
A brief history of scattering theory is given in Section 4.2 of Ref.~[\onlinecite{Benenti2017Jun}], including both electrical,\cite{Landauer1970Apr,Engquist1981Jul,Buttiker1986Oct} 
thermal\cite{Engquist1981Jul,Pendry1983Jul} 
and thermoelectric transport.\cite{Sivan1986Jan,Butcher1990}

Scattering theory starts by dividing the system into a small scattering region that is connected to  macroscopic reservoirs of free electrons. It is assumed that each electron traverses the scattering region from one reservoir to another without exchanging energy with other particles. In other words, when an electron enters the scattering region from a reservoir with energy $\epsilon$, it behaves as a wave with energy $\epsilon$ until it escapes into another reservoir.
Then the particle and heat flows are determined by the transmission function
$\mathcal{T}(\epsilon)$, which corresponds to the probability that an electron leaving one reservoir with energy $\epsilon$ will be transmitted to the other reservoir.
It predicts that the particle current leaving reservoir $L$ is 
\begin{equation}
I_{\rm L}=\frac{1}{h} \int_{-\infty}^{\infty} \mathrm{d} \epsilon \ \mathcal{T}(\epsilon)\left[f_{\rm L}(\epsilon)-f_{\rm R}(\epsilon)\right],
\label{eq:IL}
\end{equation}
where the electrical current is $e^{\text{-}}I_{\rm L}$ for electronic charge $e^{\text{-}}$.
Here $f_{\rm L}(\epsilon) = 1\big/\big(1+\exp\big[(\epsilon-\mu_i)/(k_{\rm B}T_i)]\big)$ is the Fermi function of reservoir $i$, with  electrochemical potential $\mu_i$ and temperature $T_i$.
Then the power generated by the thermoelectric is the 
difference in chemical potential multiplied by the particle current
\begin{equation}
P_{\rm gen} = (\mu_{\rm R}-\mu_{\rm L})\,  I_{\rm L}\,,
\label{eq:Pgen}
\end{equation}
with negative $P_{\rm gen}$ meaning power is absorbed and turned into heat via the process of Joule heating.
The heat current leaving reservoir L is 
\begin{equation}
J_{\rm L}=\frac{1}{h} \int_{-\infty}^{\infty} \mathrm{d} \epsilon \ \left(\epsilon-\mu_{\rm L}\right)\mathcal{T}(\epsilon)\left[f_{\rm L}(\epsilon)-f_{\rm R}(\epsilon)\right].
\label{eq:JL}
\end{equation}
Without loss of generality, we take $\mu_{\rm L}=0$, this corresponds measuring all energies from the electrochemical potential of reservoir L (i.e., defining $\eps=0$ to be at that electrochemical potential).

For a heat engine, the efficiency is defined
as the power generated over the heat flow out of the hotter reservoir. Let us
take the left (L) reservoir to be hotter, then the heat-engine efficiency is 
\begin{eqnarray}
\eta_{\rm eng} \equiv \frac{P_{\rm gen}}{J_{\rm L}} = 
\frac{\int_{-\infty}^{\infty} \mathrm{d} \epsilon \ \mu_{\rm R} \ \mathcal{T}(\epsilon)\left[f_{\rm L}(\epsilon)-f_{\rm R}(\epsilon)\right]}
{\int_{-\infty}^{\infty} \mathrm{d} \epsilon \ \epsilon \ \mathcal{T}(\epsilon)\left[f_{\rm L}(\epsilon)-f_{\rm R}(\epsilon)\right]}\,,
\label{eq:eta-eng}
\end{eqnarray}
with subscript ``eng'' for engine. The same theory has recently been applied to model hot-carrier photovoltaics,\cite{tesser2023thermodynamic} 
and new multi-terminal photovoltaics.\cite{Bertin2025Improving}

For a refrigerator, the power output is its {\it cooling power}, defined as the rate at which it extracts heat from the reservoir being refrigerated.  
We assume that reservoir L is the one being cooled, so the cooling power is $J_{\rm L}$.
The refrigerator's efficiency (often called the {\it coefficient of performance})
is the cooling power divided by the power supplied.  
The power supplied is positive (corresponding to negative power generation) and so equals $-P_{\rm gen}$.
Then the refrigerator efficiency is 
\begin{eqnarray}
\eta_{\rm fri} \equiv \frac{J_{\rm L}}{-P_{\rm gen}} = 
\frac
{\int_{-\infty}^{\infty} \mathrm{d} \epsilon \ \epsilon \ \mathcal{T}(\epsilon)\left[f_{\rm L}(\epsilon)-f_{\rm R}(\epsilon)\right]}{\int_{-\infty}^{\infty} \mathrm{d} \epsilon \ (-\mu_{\rm R}) \ \mathcal{T}(\epsilon)\left[f_{\rm L}(\epsilon)-f_{\rm R}(\epsilon)\right]}\,,
\label{eq:eta-fri}
\end{eqnarray}
with subscript ``fri'' for fridge.

\subsection{Transmission functions}

For our first example, the finite-height potential barrier,
we assume that the barrier is thick enough that there is negligible tunneling through it, so electrons with less energy than the potential-barrier's height, $\epsilon_0$, are always reflected by the barrier, while those at higher energy will traverse the barrier. For quantum point contacts, a voltage applied to a gate at the point-contact can be used to change $\epsilon_0$ to any desired value.
We assume that the system supports only one transverse mode, then the transmission function is a step-function;
\begin{equation}
\mathcal{T}_{\rm barr}(\epsilon)=\left\{\begin{array}{cl}
0  & \quad \epsilon < \epsilon_0 \, ,
\\
1 & \quad \epsilon > \epsilon_0 \, .
\end{array}\right.
\label{eq:Tbarr}
\end{equation}
While this step is sharp, real barriers and point contacts have 
$\mathcal{T}_{\rm barr}(\epsilon)$ that smoothly changes from 0 to 1 over a finite energy window, due to tunneling through the barrier\cite{Buttiker1990Apr,reviewcomment} at energies close to $\epsilon_0$.  However, a sharper step in $\mathcal{T}_{\rm barr}(\epsilon)$, allows a higher thermoelectric efficiency.\cite{Kheradsoud2019Aug}  Thus, it is best to minimize tunneling by making the barrier (or quantum point contact) longer, which makes the energy window for tunneling become exponentially small. Here, we assume the barrier is long enough to make this window much narrower than the Fermi function of the coldest reservoir, then the form of Eqs.~(\ref{eq:IL}-\ref{eq:eta-fri}) means that Eq.~(\ref{eq:Tbarr}) is sufficient to model the currents and powers.

As mentioned above, there are many experiments on thermoelectricity in systems with this sort of step-transmission function \cite{Mykkanen2020Apr,Shakouri1997Sep,Shakouri1998Jan,Molenkamp1990Aug,Dzurak1993Oct,Brantut2013Nov,footnote-atoms}, and their quantum thermodynamics has been widely studied theoretically, including 
Ref.~[\onlinecite{Kheradsoud2019Aug}] which studies how the smoothing of the step in Eq.~(\ref{eq:Tbarr}) affects the power generation, efficiency, fluctuations and Thermodynamic Uncertainty Relations.
We also note 
that transmissions similar to step-functions have been proposed for thermoelectric heat engines in quantum Hall systems \cite{samuelsson2017optimal} or quantum spin Hall systems, \cite{Hajiloo2020Oct-nonlinear-cooling} and that
the theory of the mobility edge in disordered systems can give a variety of transmission functions, with some being similar to such a step function. \cite{Sivan1986Jan,maiti2013mobility,yamamoto2017thermoelectricity,chiaracane2020quasiperiodic,khomchenko2024influence,khomchenko2024corrigendum}

Here, we assume that the barrier has only one transverse mode, since a barrier  with multiple transverse modes has a different-shaped transmission, see section 4.4.1 of Ref.~[\onlinecite{Benenti2017Jun}].
To get significant power outputs,  one could engineer many such single-transverse-mode barriers in parallel between hot and cold. Then the results
given here are for the power per mode.

In experiments where the barrier is a quantum point contact, $\epsilon_0$ is tuned by changing the voltage on a gate at the point-contact.  In experiments where the barrier is made of a different material, 
$\epsilon_0$ is tuned by changing the material. The value of $\mu_{\rm R}$ is tuned by the choice of the load's resistance.
Thus, we want to find the $\epsilon_0$ and $\mu_{\rm R}$ that maximize the efficiency at a desired power output.

\begin{figure*}
\includegraphics[width=\textwidth]{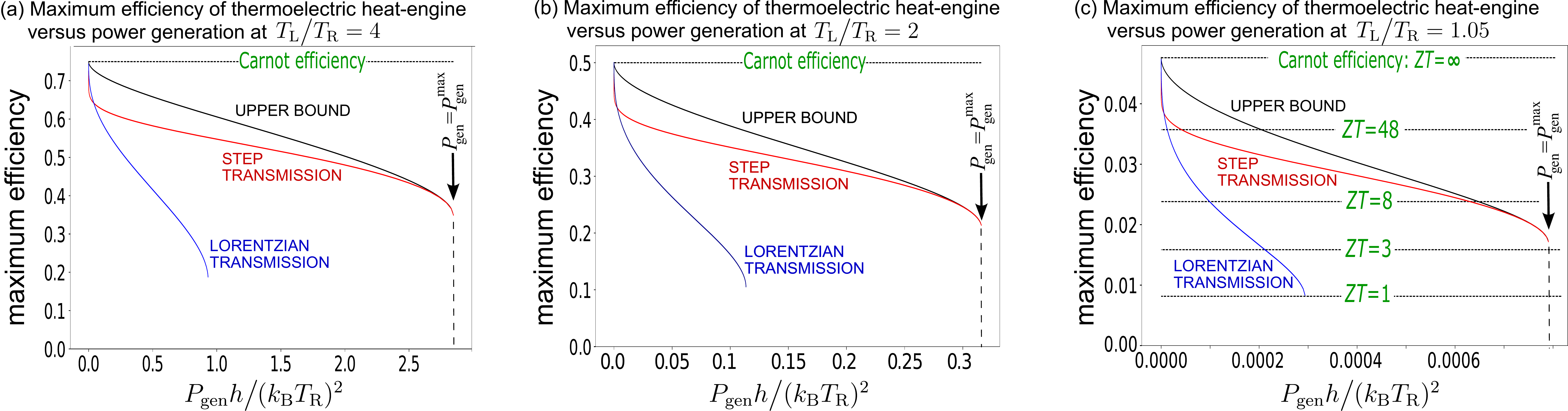} 
\caption{
Plot of the ideal heat-engine efficiency (black curve), 
the maximum efficiency achievable with a step transmission function (red curve), 
and the maximum efficiency achievable with Lorentzian transmission function (blue curve). Each one is plotted versus the power generation, $P_{\rm gen}$, for different ratios of temperature between the two reservoirs, with (a)  $T_{\rm L}/T_{\rm R}=4$, (b)  $T_{\rm L}/T_{\rm R}=2$, and (c) $T_{\rm L}/T_{\rm R}=1.05$. The horizontal black dashed lines represent the Carnot efficiency for each plot, and the arrow indicates maximum possible power generation, 
$P_{\rm gen}^{\rm max}$, see Eq.~(\ref{Eq:Pmax}). Plot (c) is in the linear response regime 
$T_{\rm L}-T_{\rm R} \ll T_{\rm R}$, for which there is a one-to-one correspondence between 
the efficiency and the dimensionless thermoelectric figure of merit called $ZT$ (see e.g., Table~1 in  Ref.~[\onlinecite{Benenti2017Jun}]), so we superimpose lines of $ZT$ in that plot.
\label{fig:best-engine}
}
\end{figure*}

Our second example corresponds to a double-barrier structure or a quantum dot\cite{prance2009electronic,Fahlvik2012Mar,Josefsson2018Oct} or to a molecular junction,\cite{Reddy2007Mar,Dubi2011Mar,cui2017perspective,sowa2019marcus,kirchberg2022energy,cohen2025unusually} in which transport occurs through a single level at energy $\epsilon_0$. For this, we take the transmission function 
\begin{equation}
\mathcal{T}_{\rm dot}(\epsilon) =\frac{\Gamma^2}{\left(\epsilon-\epsilon_0\right)^2+\Gamma^2}\,.
\label{eq:Tdot}
\end{equation}
The level's energy $\epsilon_0$ is tuned by changing the inter-barrier distance for a double-barrier structure, or by adjusting the dot size for a quantum dot. In either case, the level's coupling to reservoirs, $\Gamma$, is tuned by changing the thickness of the tunnel barrier between the single level and the reservoirs.  For molecules, it is the choice of molecule that determines $\epsilon_0$ and $\Gamma$. Similarly, the value of $\mu_{\rm R}$ is tuned by the choice of the load's resistance.
Thus, we want to find the $\epsilon_0$, $\Gamma$ and $\mu_{\rm R}$ that maximize the efficiency at a desired power output.
In the case of the double-barrier, we assume that the transverse dimension of the 
system is small enough for transport to be through a single level.
Once again, to get significant power outputs, one can imagine engineering many such single-level systems in parallel between hot and cold \cite{Jordan2013Feb,Sothmann2013Sep}. Then the results
given here are for the power per single-level system, and should be multiplied by the number of systems in parallel.

\section{Maximizing efficiency under the constraint of fixed power output}

Our goal here is to design the thermoelectric to maximize the efficiency for given power output, while working at given temperatures $T_{\rm L} $ and $T_{\rm R}$, since we assume that $T_{\rm L}$ and $T_{\rm R}$ are external conditions that are fixed by the context of the application.\footnote{Maximum efficiencies always grow as $T_{\rm L}/T_{\rm R}$ increases, so it is assumed that $T_{\rm L}/T_{\rm R}$ takes the largest value that is achievable in the context.}  
For example, to generate electricity from the waste heat in a car exhaust, then $T_{\rm L}$ would be the temperature of the exhaust pipe (perhaps 600\,Kelvin), and $T_{\rm R}$ would be the exterior air temperature (perhaps 300\,Kelvin). Then our goal is to maximize the efficiency for given electrical power output.  Alternatively, for refrigeration, $T_{\rm L}$ would be the temperature that we want the refrigerator to maintain when the ambient temperature is $T_{\rm R}$.  Then our goal is to maximize the refrigeration efficiency for given cooling power.

\subsection{Heat engine at fixed power generation}
\begin{figure}
\includegraphics[width=\columnwidth]{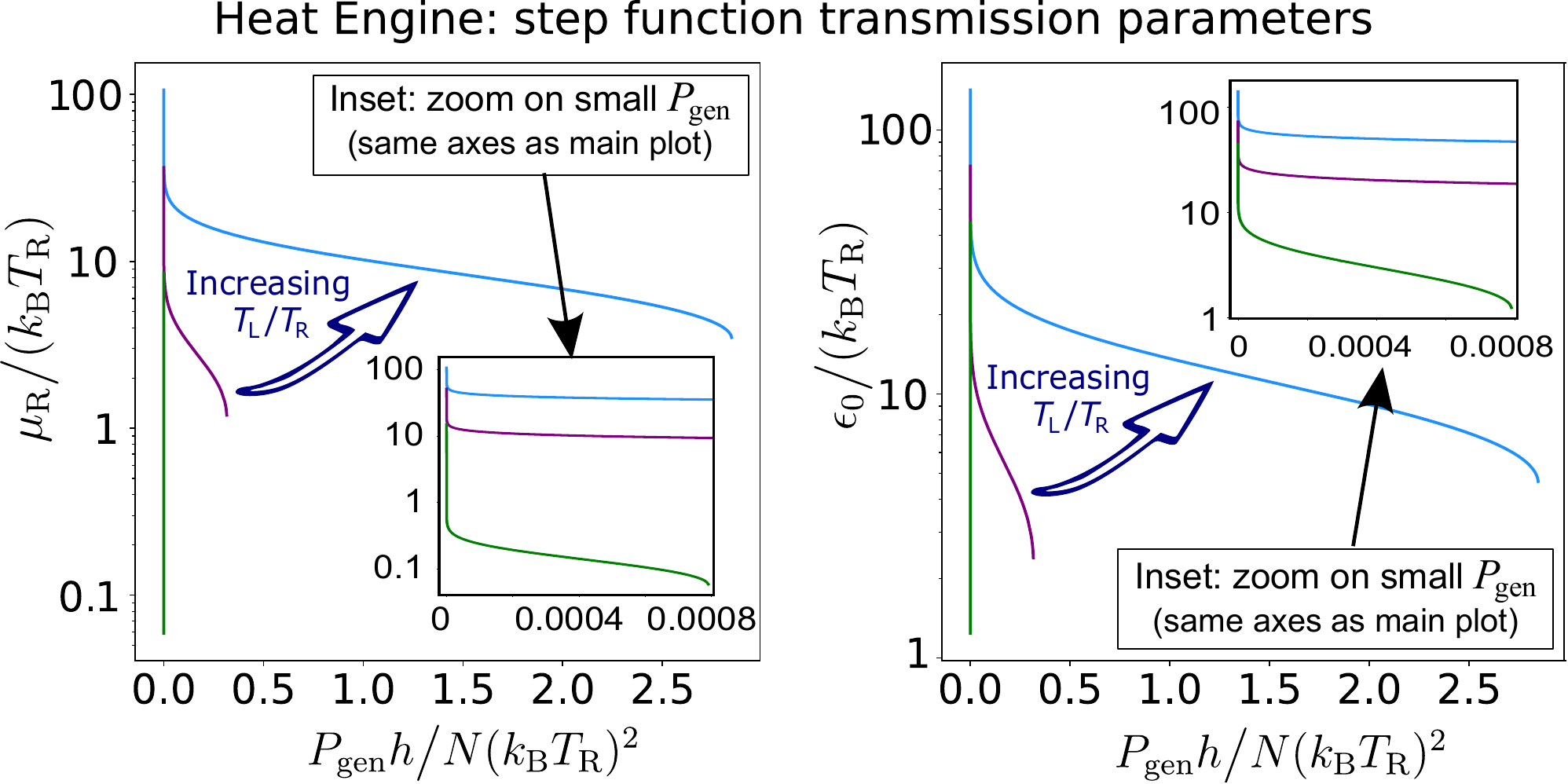} 
\caption{Plots of the step function parameters $\mu_\text{R}$ and $\epsilon_{0}$ that give
the highest heat-engine efficiency for given power generation, $P_{\rm gen}$. These parameters are plotted for $T_{\rm L}/T_{\rm R}$ equal to 4, 2 and 1.05, 
corresponding to the three red curves in Fig.~\ref{fig:best-engine}.}
\label{fig:best-engine-parameters}
\end{figure}

Let us consider using the thermoelectric as a heat engine.
Our goal is to find the parameters that maximize the efficiency
$\eta_{\rm eng}$ in Eq.~(\ref{eq:eta-eng}) under the constraint that the power generation, $P_{\rm gen}$ in Eqs.~(\ref{eq:IL},\ref{eq:Pgen}) equals a fixed value, $P$, when the transmission function is a step-function or a Lorentzian. From the definition of  
$\eta_{\rm eng}$ in Eq.~(\ref{eq:eta-eng}), we immediately see that this is the same as minimizing the heat flow $J_{\rm L}$ under the constraint that the power generation $P_{\rm gen}=P$. This is done using the Lagrange multiplier technique, which can be used to maximize or minimize (in our case minimize) $J_{\rm L}$. As usual in the Lagrange multiplier technique,
we define a Lagrangian function that is the sum of two terms, 
the first term is the quantity we want to minimize and the second term is the Lagrange multiplier, $\lambda$, multiplied by a term that vanishes when the constraint is fulfilled.
If $J_{\rm L}$ (the function we want to minimize) depends on $n$ parameters $\{x_1,\cdots\!,x_n\}$, then the Lagrangian takes the form
\begin{eqnarray}
{\cal L}_{\rm eng}(x_1,\cdots\!,x_n,\lambda) &=& J_{\rm L}(x_1,\cdots\!,x_n) 
\nonumber \\
& & \ \  + \lambda \big[P-P_{\rm gen}(x_1,\cdots\!,x_n) \big].\ \ 
\label{eq:Leng}
\end{eqnarray}
For the step transmission $n=2$ with 
$$\{x_1,\cdots\!,x_n\} \equiv \{\epsilon_0, \mu_{\rm R}\},$$ 
and 
for the Lorentzian transmission $n=3$ with 
$$
\{x_1,\cdots\!,x_n\} \equiv \{\epsilon_0,\Gamma, \mu_{\rm R}\}.
$$ 
Then to find the values of $\{x_1,\cdots\!,x_n\}$ that correspond to stationary points (maxima, minima, etc) of $J_{\rm L}(x_1,\cdots\!,x_n)$ under the constraint $P_{\rm gen}(x_1,\cdots\!,x_n)=P$, we must solve the following $(n+1)$ coupled equations
\begin{eqnarray}
\frac{d {\cal L}_{\rm eng}(x_1,\cdots\!,x_n,\!\lambda)}{d x_i} &=& 0 \mbox{ for all $i$ from 1 to $n$}, 
\nonumber \\
\mbox{ and } \frac{d {\cal L}_{\rm eng}(x_1,\cdots\!,x_n,\!\lambda)}{d \lambda} &=&0.
\label{Eq:differentials_of_Leng}
\end{eqnarray}
Solutions of these equations can be found numerically; to ensure that this runs smoothly, we use a trick in appendix~\ref{sec:numerics}. After which one must check which correspond to minima (rather than maxima or other stationary points) of $J_{\rm L}$ under the constraint $P_{\rm gen}=P$. The results are shown in Figs.~\ref{fig:best-engine} and \ref{fig:best-engine-parameters}.

Firstly, Fig.~\ref{fig:best-engine} shows that the efficiency of the optimal step-function transmission (red curve) is close to the fundamental upper-bound on efficiency for given power generation (given by the optimal boxcar transmission).
The two curves are close at all power generation, and converge as one approaches the quantum upper-bound on power generation,\cite{Whitney2014Apr,Whitney2015Mar}
\begin{eqnarray}
P_{\rm gen}^{\rm max} =  A_0 \frac{\pi^2 k_{\rm B}^2}{h}\left(T_{\rm L}^2-T_{\rm R}^2\right),
\label{Eq:Pmax}
\end{eqnarray}
where $A_0\simeq0.0321$ (note that $P_{\rm gen}^{\rm max}$ is called $P_{\rm gen}^{\rm qb2}$ in Refs.~[\onlinecite{Whitney2014Apr,Whitney2015Mar}]).
However, we note that this is for a sharp-step of the type in Eq.~(\ref{eq:Tbarr}), if the step is rounded (due to tunneling through the barrier or reflection above the barrier) then the efficiency will be lower for any given power generation, see Ref.~[\onlinecite{Kheradsoud2019Aug}]. 

Secondly, Fig.~\ref{fig:best-engine} shows that the optimal Lorentzian transmission (blue curve) has an efficiency well below the upper-bound, except at extremely small power generation. 
Typically, the Lorentzian transmission is only out-performs the step transmission, 
when the desired cooling power output is less than about 1\% of the maximum cooling power; even then it only slightly out-performs the step transmission.

\subsection{Side comment on the efficiency's lower bound}
We have to be careful when solving Eqs.~(\ref{Eq:differentials_of_Leng}) for each $P$, to make the plots of maximum efficiency in Fig.~\ref{fig:best-engine}, 
because those equations actually have {\it two solutions} for each $P$ up to $P_{\rm gen}^{\rm max}$. One solution gives the maximum value of $J_{\rm L}$, and the other gives the minimum value of $J_{\rm L}$. We are interested in maximum efficiency $\eta_{\rm eng}$, which corresponds to the solution that minimizes $J_{\rm L}$,  
Thus, we always take the solution with lower $J_{\rm L}$.

However, it is also worth mentioning the solution of Eqs.~(\ref{Eq:differentials_of_Leng}) with higher $J_{\rm L}$, which correspond to a maximum value of $J_{\rm L}$ and gives a finite {\it lower} bound on efficiency for given power generation. While this lower bound is of little practical interest, the existence of a lower bound on efficiency (the fact that the efficiency cannot be lower than a certain {\it finite} value) can seem surprising. 
However, it is direct consequence of $P_{\rm gen} =J_{\rm L}\times \eta_{\rm eng}$, 
and there being an upper bound\cite{Pendry1983Jul} on $J_{\rm L}$. Hence, one cannot achieve a given $P_{\rm gen}$ unless $\eta_{\rm eng}$ exceeds a minimum value.  Solving 
Eqs.~(\ref{Eq:differentials_of_Leng}) gives both the minimum and maximum of $J_{\rm L}$ for given power generation. When performing a numerical search for maximum efficiency (minimum of $J_{\rm L}$) from the solutions of Eqs.~(\ref{Eq:differentials_of_Leng}), one often find the wrong solution, corresponding to the minimum efficiency (maximum of $J_{\rm L}$), rather than desired solution corresponding to maximum efficiency (minimum of $J_{\rm L}$).
When this happens, one must change the numerical method's initial seed parameters
until one arrives at the desired solution.  We typically do this by finding the parameters for the desired solution at given power generation, and the use those parameters as a seed to find the desired solution at a slightly different power generation, repeating until we have those parameters at all power generations.
As the minimum efficiency is of little practical interest, we only show the maximum efficiency in our plots.

\subsection{Refrigerator efficiency at fixed cooling power}
\label{sec:fridge}

\begin{figure*}
\includegraphics[width=\textwidth]{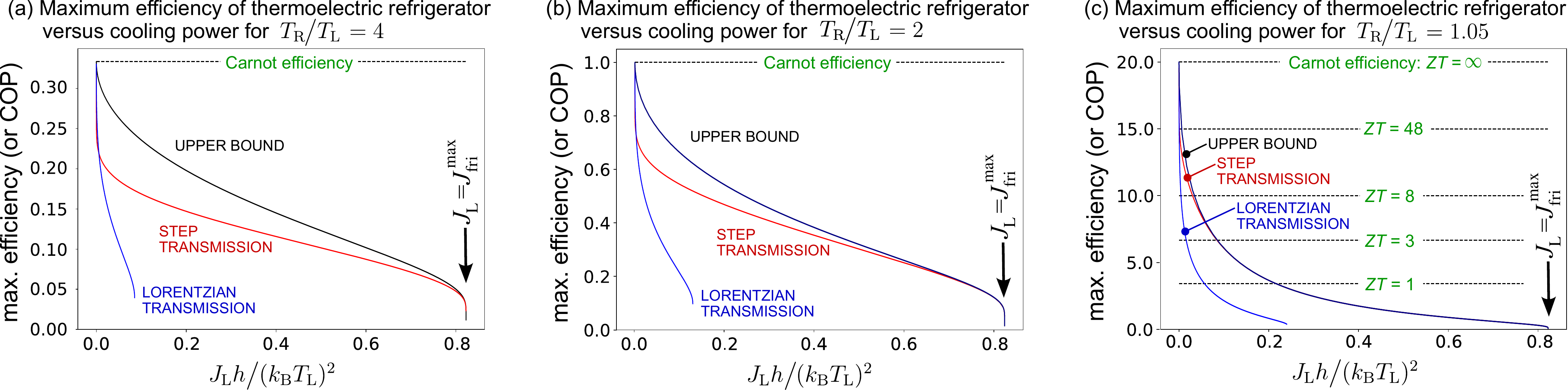}  
\caption{
Plot of the ideal refrigerator efficiency (black curve), 
the maximum efficiency achievable with a step transmission function (red curve), 
and the maximum efficiency achievable with Lorentzian transmission function (blue curve). Each one is plotted versus the cooling power required, for different ratios of temperature between the two reservoirs, with (a)  ${T_{\rm R}}/{T_{\rm L}}=4$, (b)  ${T_{\rm R}}/{T_{\rm L}}=2$, (c) ${T_R}/{T_{\rm L}}=1.05$. The horizontal black dashed lines represent the Carnot efficiency for each plot, and the arrows indicate the maximum possible cooling power, $J_{\rm fri}^{\rm max}$, see Eq.~(\ref{Eq:Jmax}).  For the plot with $T_{\rm R}/T_{\rm L} =1.05$, one is in the linear response regime, where each efficiency corresponds to a given $ZT$, hence we indicate some values of $ZT$ with green horizontal lines. }
\label{fig:best-fridge}
\end{figure*}

When we consider using the thermoelectric as a refrigerator (Peltier cooler),
our goal is to find the parameters of $\mathcal{T}(\epsilon)$ that maximize the refrigeration efficiency $\eta_{\rm fri}$ in Eq.~(\ref{eq:eta-fri})
under the constraint that the cooling power, $J_{\rm L}$ in Eqs.~(\ref{eq:JL}) equals a fixed value, $J$.
From the definition of  $\eta_{\rm fri}$ in Eq.~(\ref{eq:eta-fri}), we immediately see that this is the same as minimizing the power consumption $-P_{\rm gen}$ under the constraint that the heat flow $J_{\rm L}=J$.
This is done using the Lagrange multiplier technique
by defining
\begin{eqnarray}
{\cal L}_{\rm fri}(x_1,\cdots\!,x_n,\lambda) &=& -P_{\rm gen}(x_1,\cdots\!,x_n) 
\nonumber \\
& & \ \  + \lambda \big[J-J_{\rm L}(x_1,\cdots\!,x_n) \big].\ \ 
\label{eq:Lfri}
\end{eqnarray}
We then proceed with the usual Lagrange multiplier technique, just as for the heat engine above. The results are shown in Figs.~\ref{fig:best-fridge}.

The first thing that we see in Fig.~\ref{fig:best-fridge} is that the efficiency of the optimal step-function transmission (red curve) is close to the fundamental upper-bound on efficiency for given cooling power (given by the optimal boxcar transmission).
The two curves are close at all power generation, and converge as one approaches the quantum upper-bound on cooling power,\cite{Whitney2014Apr,Whitney2015Mar}
\begin{eqnarray}
J_{\rm fri}^{\rm max} =  \frac{\pi^2}{12h} k_{\rm B}^2 T_{\rm L}^2,
\label{Eq:Jmax}
\end{eqnarray}
which is half Pendry's upper bound on heat-flow in a single mode\cite{Pendry1983Jul} (called $J_{\rm L}^{\rm qb}$ in Refs.~[\onlinecite{Whitney2014Apr,Whitney2015Mar}]).
The second thing that we see in Fig.~\ref{fig:best-engine} is that the efficiency of the optimal Lorentzian transmission (blue curve) is far from optimal, except at extremely small cooling power.
Typically, the Lorentzian transmission is only out-performs the step transmission, 
when the desired cooling power output is less than about 1\% of the maximum cooling power;  even then it only slightly out-performs the step transmission.

\section{Analytical treatment for the finite-height barrier}

For the finite height barrier with transmission given by the step function in Eq.~(\ref{eq:Tbarr}),
we can perform the integrals in Eqs.~(\ref{eq:IL},\ref{eq:JL}) to get 
the following
analytic forms for the currents and power outputs:
\begin{eqnarray}
J_{\rm L}\left(\epsilon_0, \mu_{\rm R}\right) &=& F_{\rm L}\left(\epsilon_0\right)-F_{\rm R}\left(\epsilon_0, \mu_{\rm R}\right), 
\label{eq:JL-step}\\
P_{\mathrm{gen}}\left(\epsilon_0, \mu_{\rm R}\right) &=& \mu_{\rm R}\left[G_{\rm L}\left(\epsilon_0\right)-G_{\rm R}\left(\epsilon_0, \mu_{\rm R}\right)\right],
\label{eq:Pgen-step}
\end{eqnarray}
where we define for compactness, 
\begin{eqnarray}
F_{\rm L}(\epsilon_{0}) &=&\epsilon_{0} G_{\rm L}(\epsilon_{0})-\frac{\left(k_{\mathrm{B}} T_{\rm L}\right)^2}{h} \operatorname{Li}_2\left[-\mathrm{e}^{-\epsilon_{0} /\left(k_{\mathrm{B}} T_{\rm L}\right)}\right]\!, \qquad
\label{eq:FL}
\\
G_{\rm L}(\epsilon_{0}) &=& \frac{k_{\mathrm{B}} T_{\rm L}}{h} \ln \left[1+\mathrm{e}^{-\epsilon_{0} /\left(k_{\mathrm{B}} T_{\rm L}\right)}\right], 
\end{eqnarray}
\begin{eqnarray}
F_{\rm R}(\epsilon_{0},\mu_{\rm R}) &=&\epsilon_{0} G_{\rm R}(\epsilon_{0},\mu_{\rm R})
\nonumber 
\\
& &-\frac{\left(k_{\mathrm{B}} T_{\rm R}\right)^2}{h} \operatorname{Li}_2\left[-\mathrm{e}^{-\left(\epsilon_{0}-\mu_{\rm R}\right) /\left(k_{\mathrm{B}} T_{\rm R}\right)}\right]\!, \qquad
\\
G_{\rm R}(\epsilon_{0},\mu_{\rm R}) &=&\frac{k_{\mathrm{B}} T_{\rm R}}{h} \ln \left[1+\mathrm{e}^{-\left(\epsilon_{0}-\mu_{\rm R}\right) /\left(k_{\mathrm{B}} T_{\rm R}\right)}\right].
\label{eq:GR}
\end{eqnarray}
Here, $
\mathrm{Li}_2(x)=-\int_0^x \frac{\ln (1-t)}{t} d t$ is the dilogarithm function.
Then for the heat engine, Eq.~(\ref{eq:Leng}) reduces to
\begin{eqnarray}
{\cal L}_{\rm eng}\left(\epsilon_0, \mu_{\rm R}, \lambda\right)
&=&J_{\rm L}\left(\epsilon_0, \mu_{\rm R}\right)
+\lambda \left[P-P_{\rm gen}\left(\varepsilon_0, \mu_{\rm R}\right)\right],\quad
\nonumber \\
\end{eqnarray}
where $P$ is the desired power generation.
Taking the derivative with respect to $\eps_0$, $\mu_{\rm R}$ and $\lambda$
gives the following three nonlinear simultaneous equations to solve,
\begin{eqnarray}
0 \!=&\! \frac{d J_{\rm L}}{d\eps_0} - \lambda \frac{d P_{\rm gen}}{d\eps_0} \,= &
F'_{\rm L} - F'_{\rm R} - \lambda \mu_{\rm R} \left(G'_{\rm L} - G'_{\rm R}\right),
\qquad 
\label{Eq:simultaneous1}\\
0 \!=&\! \frac{d J_{\rm L}}{d\mu_{\rm R}} - \lambda \frac{d P_{\rm gen}}{d\mu_{\rm R}}
\,=& F'_{\rm R} -G_{\rm R} 
\nonumber \\
& & - \lambda \left(G_{\rm L} - G_{\rm R} +\mu_{\rm R} G'_{\rm R}\right),
\qquad 
\label{Eq:simultaneous2}\\
0 \!=&\! P-P_{\rm gen} \ \ \ =&\!\! P-\mu_{\rm R}\left(G_{\rm L} - G_{\rm R}\right),
\label{Eq:simultaneous3}
\end{eqnarray}
where for compactness we drop the arguments from $G_i$ and $F_i$, 
and use the prime to indicate a derivative with respect to $\eps_0$.
We also use the fact that 
$\big({\rm d}G_{\rm R}\big/{\rm d}\mu_{\rm R}\big)= -\big({\rm d}G_{\rm R}\big/{\rm d}\eps_0\big)$ and
$\big({\rm d}F_{\rm R}\big/{\rm d}\mu_{\rm R}\big)= -\big({\rm d}F_{\rm R}\big/{\rm d}\eps_0\big)+ G_{\rm R}$, 
while 
$\big({\rm d}G_{\rm L}\big/{\rm d}\mu_{\rm R}\big)= \big({\rm d}F_{\rm L}\big/{\rm d}\mu_{\rm R}\big) = 0$.
Rearranging Eqs.~(\ref{Eq:simultaneous1},\ref{Eq:simultaneous2}) to get two formulas for $\lambda$, equating the two and multiplying through to eliminate the denominators, we get
\begin{eqnarray}
& & \big(F'_{\rm L} - F'_{\rm R}\big) \, \left(G_{\rm L} - G_{\rm R} +\mu_{\rm R} G'_{\rm R}\right) 
\nonumber \\
& & \hskip 2cm =\, \mu_{\rm R} \left(F'_{\rm R}-G_{\rm R}\right) \left(G'_{\rm L} - G'_{\rm R}\right).
\nonumber
\end{eqnarray}
Now we use the fact that $G'_i= -f_i (\eps_0)/h$ and $F'_i= -\eps_0 f_i (\eps_0)/h$ for $i\in\{{\rm L},{\rm R}\}$, and use Eq.~(\ref{Eq:simultaneous3}) to replace $G_{\rm L} - G_{\rm R}$ with $P/\mu_{\rm R}$. Then we find that this equation 
reduces to simplify
\begin{eqnarray}
\big(f_{\rm L}(\eps_0) - f_{\rm R}(\eps_0)\big) \big[\eps_0P +\mu_{\rm R}^2G_R(\eps_0)\big] &=& 0\, . 
\label{Eq:solution1}
\end{eqnarray}
For the heat engine, we are interested in cases where $P$ and $\eps_0$ are positive (while $\mu_{\rm R}^2$ and $G_R(\eps_0)$ are always positive). Hence, the solution of interest for a heat engine 
is $f_{\rm L}(\eps_0) = f_{\rm R}(\eps_0)$, hence 
\begin{eqnarray}
\mu_{\rm R} = \left(1 - T_{\rm R}\big/T_{\rm L} \right)\eps_0\, ,
\label{Eq:optimal_muR}
\end{eqnarray}
which means that 
$G_{\rm R} = \big(T_{\rm R}\big/T_{\rm L}\big) G_{\rm L}$.
Hence, we find that $\eps_0$ for given $P$ is the solution to the equation
\begin{eqnarray}
\frac{\eps_0}{\kB T_{\rm L}} \ln \left[1+\exp \left( -\frac{\eps_0}{\kB T_{\rm L}}\right) \right] 
&=& \frac{\hbar P}{\kB^2(T_{\rm L} -T_{\rm R})^2}\,. \qquad 
\label{Eq:optimal_eps0}
\end{eqnarray}
This is a complete solution to the problem for a heat engine.  For any desired power generation, $P$,
we can use Eq.~(\ref{Eq:optimal_eps0}) 
to give the optimal $\eps_0$ and then Eq.~(\ref{Eq:optimal_muR}) to give the optimal $\mu_{\rm R}$.  
Inserting these optimal values of $\eps_0$ and $\mu_{\rm R}$ into Eqs.~(\ref{eq:JL-step}-\ref{eq:GR}), then gives the maximal efficiency that step-function transmission can achieve under the constraint that the power generated, $P_{\rm gen}(\epsilon_0, \mu_{\rm R})=P$.
As the equation for  $\eps_0$ has no algebraic solution we cannot give a formula for the maximum efficiency, but the results correspond to those in Fig.~\ref{fig:best-engine}.

Unfortunately, the refrigerator is less simple; we can write an equation similar to Eq.~(\ref{Eq:solution1}), but the relevant solution (one with $P<0$) comes from the equivalent of the second term in Eq.~(\ref{Eq:solution1}). The result is that Eq.~(\ref{Eq:optimal_muR}) is replaced by an implicit equation for $\mu_{\rm R}$ as a function of $\eps_0$ and cooling power. Formally, this gives a complete algebraic solution, but can only be explored numerically, where it gives results corresponding to those  already plotted in Fig.~\ref{fig:best-fridge}.

\subsection{Optimal heat-engine efficiency when \\ power generation is fixed and small}
When we constrain the heat engine's power generation $P_{\rm gen}=P$ to be small, so 
$P\ll\kB\left(T_\text{L}^2-T_\text{R}^2\right)\big/h$, 
then Eq.~(\ref{Eq:optimal_eps0}) indicates that optimal $\epsilon_0 \gg T_{\rm L},T_{\rm R}$. 
Then Eq.~(\ref{Eq:optimal_eps0}) reduces to 
\begin{eqnarray}
\frac{\eps_0}{\kB T_{\rm L}} \exp \left( -\frac{\eps_0}{\kB T_{\rm L}}\right)
&=& \frac{\hbar P}{\kB^2\big(T_{\rm L}^2 -T_{\rm R}^2\big)}\,. \qquad 
\end{eqnarray}
In this limit, Eqs.~(\ref{eq:FL},\ref{eq:GR}) reduce to 
\begin{eqnarray}
F_{\rm L}(\epsilon_{0}) &=& \frac{\kB T_{\rm L}}{h}\left(\epsilon_{0}+\kB T_{\rm L}\right) \mathrm{e}^{-\epsilon_{0} /\left(\kB T_{\rm L}\right)},
\\
G_{\rm L}(\epsilon_{0}) &=& \frac{\kB T_{\rm L}}{h} \mathrm{e}^{-\epsilon_{0} /\left(\kB T_{\rm L}\right)},
\\
F_{\rm R}(\epsilon_{0},\mu_{\rm R}) &=& \frac{\kB T_{\rm R}}{h}\left(\epsilon_{0}+\kB T_{\rm R}\right) \mathrm{e}^{-(\epsilon_{0}-\mu_{\rm R}) /\left(\kB T_{\rm R}\right)}, \quad \ 
\\
G_{\rm R}(\epsilon_{0},\mu_{\rm R}) &=& \frac{\kB T_{\rm R}}{h} \mathrm{e}^{-(\epsilon_{0}-\mu_{\rm R}) /\left(\kB T_{\rm R}\right)}.
\end{eqnarray}
Hence, recalling Eq.~(\ref{Eq:optimal_muR}), this limit gives 
\begin{eqnarray}
P_{\rm gen} &=& \frac{\kB^2}{h} (T_{\rm L}-T_{\rm R})^2\,\frac{\epsilon_0}{\kB T_{\rm L}}
\mathrm{e}^{-\epsilon_{0} /\left(\kB T_{\rm L}\right)} ,
\label{Eq:Pgen-smallP-1}
\\
J_{\rm L} &=& \frac{\kB}{h}
\big(T_{\rm L}-T_{\rm R}\big)\big(\epsilon_0 +\kB (T_{\rm L}+T_{\rm R}) \big) \mathrm{e}^{-\epsilon_{0} /\left(\kB T_{\rm L}\right)}.\qquad
\label{Eq:JL-smallP-1}
\end{eqnarray}
Substituting Eq.~(\ref{Eq:Pgen-smallP-1}) into Eq.~(\ref{Eq:JL-smallP-1}) allows us to write the {\it minimum} heat flow under the constraint that $P_{\rm gen}=P$, for small $P$  
\begin{eqnarray}
J_{\rm L}(P) &=& \frac{T_{\rm L}}{T_{\rm L}-T_{\rm R}}\left( 1 + \frac{\kB (T_{\rm L}+T_{\rm R})}{\eps_0(P)} \right) \, , \qquad
\label{Eq:JL-smallP-2}
\end{eqnarray}
where $\eps_0(P)$ is the optimal barrier height for $P_{\rm gen}=P$, found by inverting Eq.~(\ref{Eq:Pgen-smallP-1}) to get $\eps_0$ in terms of $P_{\rm gen}$.  This inversion 
uses the result in Appendix~\ref{app:inverting}, giving
\begin{eqnarray}
\frac{\eps_0(P)}{k_{\rm B}T_{\rm L}} =  \ln (P_0/P)  
+ \!\left(\!1\!+\! \frac{1}{\ln(P_0/P)}\!\right)  \ln\!\big(\ln(P_0/P)\big),\ \
\label{Eq:eps0-smallP}
\end{eqnarray}
for $P\ll P_0$, with
$P_0= \frac{1}{h}k_{\rm B}^2\big(T_{\rm L}-T_{\rm R}\big)^2$.

When we write small $P$, we mean $P\ll P_0$. However, we also
note that $P_0$ is naturally written in terms of 
the quantum upper-bound on power generation in Eq.~(\ref{Eq:Pmax}), 
as follows
\begin{eqnarray}
P_0= \frac{1}{A_0} \, \frac{T_{\rm L}-T_{\rm R}}{T_{\rm L}+T_{\rm R}}\ P_{\rm gen}^{\rm max}\, ,
\end{eqnarray}
so $P\ll P_0$ is equivalent to $P \ll P_{\rm gen}^{\rm max}$ for all finite  values of the temperature difference, $(T_{\rm L}-T_{\rm R})$.

Hence, the maximum efficiency for a step-function transmission at small power generation, $P$ is given by 
\begin{eqnarray}
\eta_{\rm eng}(P \ll P_{\rm gen}^{\rm max}) \, =\,   \eta_{\rm eng}^{\rm Carnot} \frac{\eps_0(P)}{\eps_0(P)\!+\!k_{\rm B}(T_{\rm L}+T_{\rm R})}\,,\ 
\end{eqnarray}
where $\eps_0(P)$ is the optimal barrier height for the desired $P$ given by Eq.~(\ref{Eq:eps0-smallP}). Since  $\eps_0(P)$ goes to $\infty$ as $P\to 0$, we see that this efficiency goes to the Carnot efficiency at vanishing power generation.
However, the maximum efficiency is less than Carnot efficiency for finite $P$. 

\subsection{Difference in parameters for step-function versus boxcar}

The difference between the step-function and the boxcar transmission function (which Refs.~[\onlinecite{Whitney2014Apr,Whitney2015Mar}] showed can achieve the ideal efficiency at given power), is that the boxcar also blocks the flow of electrons at high energies. Thus, one might guess that high-energy electrons are of little importance, and this is the origin of the step-function being close to optimal. Here we show that this guess is right at high power outputs, but completely wrong at low power outputs.

Let us start with power output close to maximum. There Refs.~[\onlinecite{Whitney2014Apr,Whitney2015Mar}] showed that large power output requires the upper-bound on the boxcar function to be at high energies. While electron flow at high energies reduces the efficiency, the electron flow is small at such high energies (since this is in the tail of the Fermi functions). Hence, allowing this flow does not significantly reduce the efficiency. 
Thus, for a power output close to maximum, replacing the optimal boxcar with a step-function makes little difference to the efficiency.

Let us now turn to small power outputs, where we see that something completely different occurs; the optimal step-function looks completely different from the ideal case of an optimal boxcar.
For the optimal boxcar at small power output, Refs.~[\onlinecite{Whitney2014Apr,Whitney2015Mar}] tells us that this boxcar is very narrow, and sits at a modest energy (of order temperature above the chemical potential), with a modest potential difference (of order the temperature difference). More precisely, for a heat engine at very small power output, 
this very narrow boxcar has (setting $\mu_{\rm L}=0$),
\begin{eqnarray}
\eps_0^{\rm boxcar}&=&3.24 k_{\rm B} T_{\rm L},
\\
\mu_{\rm R}^{\rm boxcar} &=& (1-T_{\rm R}/T_{\rm L})  \,\eps_0^{\rm boxcar},
\end{eqnarray}
see Eqs.~(29,50) of Ref.~[\onlinecite{Whitney2015Mar}]). 
For the optimal step function at small power output, the step is at very high energy (so the current is exponentially small) with a huge potential difference. More precisely, for a heat engine at very small power output,
the best step function has
\begin{eqnarray}
\eps_0^{\rm step} &\to& \infty,
\\
\mu_{\rm R}^{\rm step} &\to& \infty,
\end{eqnarray}
since they both diverge logarithmically as the power output goes to zero, as seen from Eqs.~(\ref{Eq:optimal_muR}) and (\ref{Eq:eps0-smallP}) above.

The boxcar and the step-function thereby approach Carnot efficiency at small power output
in very different ways.
The boxcar approaches Carnot efficiency as we make it narrower and narrower, but at the cost of the power generation going to zero like a power law in boxcar width.  The step-function approaches Carnot efficiency as we raise the barrier very high, but at the cost of the current (and hence the power generation) going to zero exponentially fast as we raise the barrier.
Despite this, the best step-function efficiency can get remarkably close to that of the optimal boxcar at any given power output.

\section{Phonons and heat leaks}

In real systems, the efficiency of the thermoelectric response of the electrons is not the whole story; the overall efficiency is always reduced by heat leaks.  These heat leaks include all heat-flows between hot and cold that are unrelated to the electron flow in 
the thermoelectrics.  This includes phonon flows through the thermoelectric and any insulating-material between the hot and cold reservoir, and the photon flows through any free-space between the hot and cold reservoirs. Typically, even after such heat leaks have been minimized (by thermal insulation between hot and cold, etc), they is still significant.

We quantify all such heat leaks as a heat flow $J_{\rm leak}$, which we assume is an arbitrary function of  $T_{\rm L}$ and $T_{\rm R}$, but is independent of the electron flow, and so is independent of ${\cal T}(\epsilon)$. 
This assumption that phonon and photon flows are independent of the electron flow is reasonable for two reasons.  Firstly, phonons or photons flow everywhere (not just through the thermoelectric) and they often carry a significant heat between hot and cold reservoirs through insulating material (or free space) in places around those reservoirs where the thermoelectric is absent; this flow gives a contribution to heat leaks that depends on $T_{\rm L}$ and $T_{\rm R}$, but it is clearly independent of all properties of the thermoelectric.\footnote{There is a direct analogy here with a household refrigerator. There the heat leaks are often dominated by heat flow through the insulating material around the cold compartment. This heat flow is unrelated to the properties of the refrigeration circuit.} Secondly, for phonons flowing through the thermoelectric, nanostructuring that modifies electron flows often has little effect on phonon flows. For example, as phonons are not charged particles, they are insensitive to changes in the electrochemical potentials, or changes in the height of a barrier to electrons (typically done by doping a region of semi-conductor). So phonon flows will depend on  $T_{\rm L}$ and $T_{\rm R}$, but will be insensitive to changes of parameters necessary to optimize the thermoelectric response. Instead, phonon flows are affected by things like lattice mismatches, that electrons are fairly insensitive to.  Of course, 
electron and phonon flows could be coupled by electron-phonon interactions inside the nanostructure, but our model assumes that electrons traverse the nanostructure fast enough that 
there is no time for this interaction to occur.\footnote{Scattering theory relies on this assumption; this assumption's legitimacy comes from this theory often describing experimental observations. However, there are also systems where scattering theory is inapplicable because of strong electron-phonon interactions. We mention such issues in our conclusions.}     
Hence, here we
assume this heat leak due to phonons and photons has a fixed value (given by $T_{\rm L}$, $T_{\rm R}$ and material properties), that is unchanged when we maximize the efficiency by adjusting the electron flow in the thermoelectric.

Section~XIV of Ref.~[\onlinecite{Whitney2015Mar}] showed that, in the presence of heat leaks, the highest efficiency for given power output is still given by a boxcar transmission function.  This was further investigated recently 
in the linear response regime in Ref.~[\onlinecite{Ding2023Oct}].  
Here, we study how close the step transmission and Lorentzian transmissions are to an optimal boxcar transmission in the presence of heat leaks.

\begin{figure*}
\centerline{\includegraphics[width=\textwidth]{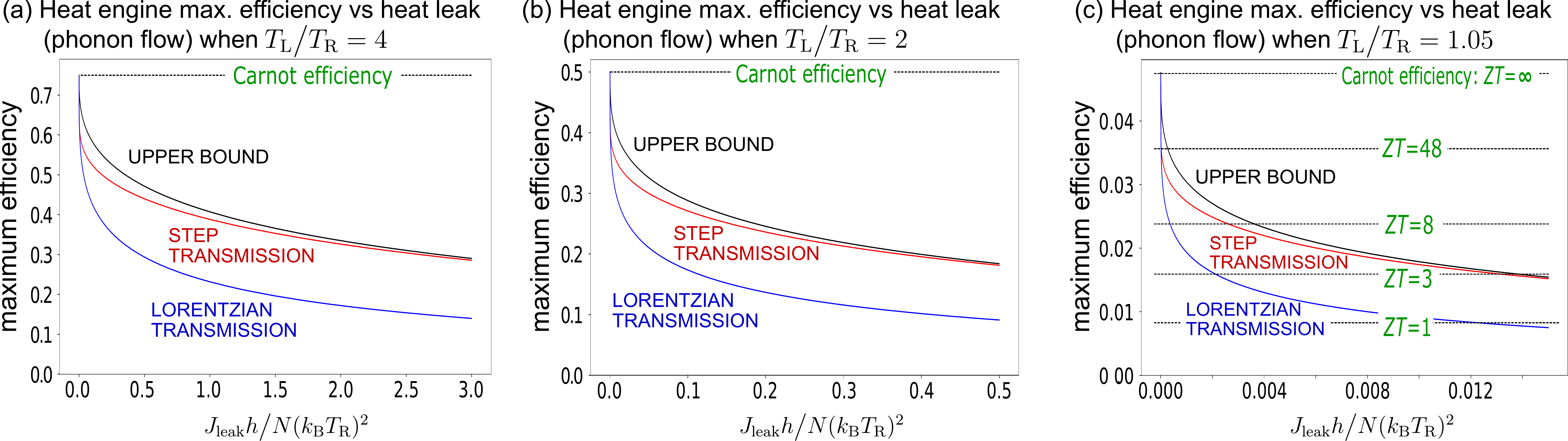}}
\caption{
Plot of the maximum heat-engine efficiency for given $J_{\rm leak}$ at various values of temperature.
The black curves are the upper bounds given by the optimal boxcar transmission, the 
red curves are for the optimal step-function transmission, and the blue curves are for the optimal Lorentzian transmission.  For each value of $J_{\rm leak}$, the efficiency is maximized over all transmission parameters and also over all values of power generation, and we plot this maximal efficiency. For the plot with $T_{\rm L}/T_{\rm R} =1.05$, one is in the linear response regime, where each efficiency corresponds to a given $ZT$, hence we indicate some values of $ZT$ with green horizontal lines. 
\label{Fig:heat-engine-leaks}
}
\end{figure*}

\subsection{Heat engine with heat leaks}
\label{sec:heat-eng-leaks}

Following Ref.~[\onlinecite{Whitney2015Mar}], we note that a heat engine's efficiency in the presence of heat leaks is
\begin{eqnarray}
\eta^{\rm with\, leak}_{\rm eng}(P) \equiv \frac{P}{J_{\rm leak}+J(P)} = \frac{P}{J_{\rm leak}+P/\eta_{\rm eng}(P)}\,,
\label{eq:eta_eng_with_leak}
\end{eqnarray}
where $\eta_{\rm eng}(P)$ is the efficiency without heat leaks, as evaluated in earlier 
sections of this article.

One can see by inspection that for given $P$ and $J_{\rm leak}$, Eq.~(\ref{eq:eta_eng_with_leak}) is maximized by maximizing $\eta_{\rm eng}(P)$. This tells us that the transmission that maximizes the efficiency for given power is independent of strength of heat leaks (phonons, photons, etc), the best transmission in the absence of heat leaks is also the best transmission in the presence of arbitrarily strong heat leaks. 

However, this is only the maximum efficiency for given power generation, $P$. This efficiency is zero at $P=0$, it grows with $P$ to a maximum at a specific value of $P$ (this value of $P$ grows with increasing $J_{\rm leak}$).
Examples of this are shown in Fig.~12a of Ref.~[\onlinecite{Whitney2015Mar}].
Here, we are interested in the best possible efficiency (without the constraint of a given power generation), we take the maximum of $\eta^{\rm with\, leak}_{\rm eng}(P)$ at each $P$, and then find the value of $P$ with the best $\eta^{\rm with\, leak}_{\rm eng}$. 
Defining $P_\text{best-$\eta$}$ as the power which maximizes $\eta^{\rm with\, leak}_{\rm eng}(P)$, it is a solution of $\big(d\eta^{\rm with\, leak}_{\rm eng}(P)\big/dP\big)\big|_{P=P_\text{best-$\eta$}}=0$.
From Eq.~(\ref{eq:eta_eng_with_leak}) we find this corresponds to
\begin{eqnarray}
\Big(\eta_{\rm eng}(P_\text{best-$\eta$})\Big)^2 = - \frac{P^2_\text{best-$\eta$}}{J_{\rm leak}} \,\left.\frac{d\eta_{\rm eng}(P)}{dP}\right|_{P=P_\text{best-$\eta$}}.
\label{eq:crossing-curves}
\end{eqnarray}
In all cases considered here, Eq.~(\ref{eq:crossing-curves})'s left-hand side is a positive monotonically-decaying function, and Eq.~(\ref{eq:crossing-curves})'s right-hand side is a monotonically growing function that is zero at $P=0$ and diverges at $P_{\rm gen}^{\rm max}$ (see Eq.~\ref{Eq:Pmax}).

Thus, these two functions on the left and right hand sides of Eq.~(\ref{eq:crossing-curves}) will always cross, and the crossing point defines the power generation $P_\text{best-$\eta$}$ that gives the best efficiency  for given $J_{\rm leak}$. When heat leaks are vanishingly small (meaning $J_{\rm leak}/P_{\rm max} \to 0$) then $P_\text{best-$\eta$} \to 0$. However, as heat leaks grow, so does $P_\text{best-$\eta$}$.
In the limit of extremely strong heat leaks (meaning $J_{\rm leak}/P_{\rm max} \to \infty$), we see that $P_\text{best-$\eta$}\to P_{\rm max}$.

Fig.~\ref{Fig:heat-engine-leaks}
shows this for a thermoelectric heat engine with three types of transmission function: the optimal boxcar-function transmission, step-function transmission, and Lorentzian transmission.
Once again, we see that the step-function transmission achieves an efficiency close to that of the optimal boxcar transmission. The Lorentzian transmission is worse than the step transmission, except at extremely small $J_{\rm leak}$, corresponding to heat leaks (due to phonons and photons) being less than about 1\% of the total heat flow; even then the Lorentzian transmission  only slightly out-performs the step transmission.

\subsection{Refrigerator with heat leaks}

Following Ref.~[\onlinecite{Whitney2015Mar}], we note that a refrigerator's efficiency in the presence of heat leak $J_{\rm leak}$ flowing into the reservoir being refrigerated
is
\begin{eqnarray}
\eta^{\rm with\, leak}_{\rm fri}(J) \equiv {J \over P(J+J_{\rm leak})}=  {J \, \eta_{\rm fri}(J+J_{\rm leak}) \over J+J_{\rm leak}} ,
\label{Eq:fri-e+ph}
\end{eqnarray}
since to achieve the extraction of heat from the cold reservoir at a rate $J$ when there is a back-flow of heat $J_{\rm leak}$ actually requires extracting heat at the rate $J+J_{\rm leak}$.  Here, $\eta_{\rm fri}(J)$ is the refrigerator efficiency in the absence of heat leaks, as discussed in Sec.~\ref{sec:fridge}.

However, Eq.~(\ref{Eq:fri-e+ph}) is only the maximum efficiency for given cooling power, $J$. This efficiency is zero at $J=0$, it grows with $J$ to a maximum at a specific value of $J$ (this value of $J$ grows with increasing $J_{\rm leak}$).
Examples of this are shown in Fig.~12b of Ref.~[\onlinecite{Whitney2015Mar}].
Here, we are interested in the best possible refrigerator efficiency (without the constraint of a given cooling power) for given $J_{\rm leak}$, which can be found in a similar manner to that for the best possible heat engine in Sec.~\ref{sec:heat-eng-leaks} above.

\begin{figure*}
\centerline{\includegraphics[width=\textwidth]{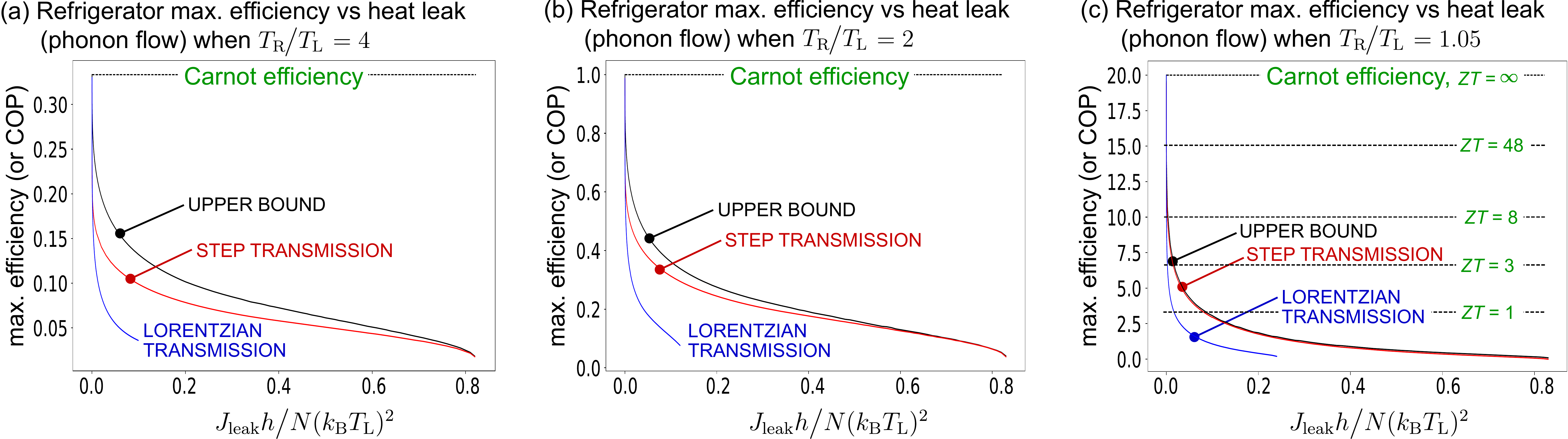}}
\caption{
Plot of the maximum refrigerator efficiency (often called the coefficient of performance or COP) for given $J_{\rm leak}$ at various values of temperature.
The black curves are the upper bounds given by the optimal boxcar transmission, the 
red curves are for the optimal step-function transmission, and the blue curves are for the optimal Lorentzian transmission.  For each value of $J_{\rm leak}$, the efficiency is maximized over all transmission parameters and also over all values of the refrigerator's cooling power, and we plot this maximal efficiency. For the plot with $T_{\rm R}/T_{\rm L} =1.05$, one is in the linear response regime, where each efficiency corresponds to a given $ZT$, hence we indicate some values of $ZT$ with green horizontal lines. 
\label{Fig:refrigerator-leaks}
}
\end{figure*}

Fig.~\ref{Fig:refrigerator-leaks} shows this for a thermoelectric heat engine with three types of transmission function: the optimal boxcar-function transmission, step-function transmission, and Lorentzian transmission.
As for the heat engine, we see that the step-function transmission achieves an efficiency close to that of the optimal boxcar transmission. The Lorentzian transmission is worse than the step transmission, except at extremely small $J_{\rm leak}$, corresponding to heat leaks (due to phonons and photons) being less than about 1\% of the total heat flow; even then the Lorentzian transmission only slightly out-performs the step transmission.

\section{Concluding remarks}

It is rare that the easiest is also the best; however, here it is almost true. 
Our modeling predicts that a quantum thermoelectric based on a simple step-function transmission will be almost as efficient as the optimal boxcar transmission function,  in most situations of interest.
Our results also show that the optimal step-function is much better than the optimal Lorentzian, except in the rare cases where the desired power output is less than about 1\% of the maximum power (see Figs. \ref{fig:best-engine} and \ref{fig:best-fridge}), and 
the heat leaks (due to phonons and photons) are less than about 1\% of the total heat flow (see Figs.~\ref{Fig:heat-engine-leaks} and \ref{Fig:refrigerator-leaks}). Even when the Lorentzian is better, it only slightly out-performs the step-function, making the step-function a good choice in all situations.

While it is worth searching for quantum thermoelectrics with the optimal boxcar transmission, we note that step-functions are about the easiest transmissions to implement experimentally, as done routinely for many years using both potential barriers and quantum point-contacts.
Of course, 
our conclusion in favor of step-function transmissions is based on a theoretical model that neglects the complexities of experimental systems, so it needs to be tested experimentally. In particular, our model is based on scattering theory, which neglects all interaction effects (both electron-electron and electron-phonon interactions) not captured by a mean-field approximation.\footnote{For example, see    
section 6.1 of the review article, Ref.~[\onlinecite{Benenti2017Jun}].}
Nonetheless, our results make it clear that step-function transmissions merit more experimental investigation.

In cases requiring a higher efficiency than that of a step-transmission,
one would have to experimentally-implement various theoretical proposals for approaching boxcar-like transmissions. Some propose using band-structure \cite{Whitney2015Mar,maiti2013mobility,yamamoto2017thermoelectricity,chiaracane2020quasiperiodic,brandner2025thermodynamic}, but we believe that the implementation could use inter-band tunneling, for which there have recently been experiments \cite{reihani2024cooling}. Others propose using Aharonov-Bohm rings \cite{haack2019efficient,haack2021nonlinear,behera2023quantum,bedkihal2025fundamental-review} to get transmission functions with a rich variety of shapes, including those similar to a boxcar function.
However, Aharonov-Bohm rings require carefully-tuned external magnetic-fields, 
which are not necessary for the optimal potential barriers or point-contacts.

For even better efficiencies, one could explore interacting systems. Simulations of certain systems with strong relaxation\cite{Brandner2015Jan} or inter-particle interactions \cite{Luo2018Aug} indicate efficiencies higher than the upper-bound established  for non-interacting electronic systems in Refs.~[\onlinecite{Whitney2014Apr,Whitney2015Mar,Whitney2016May}].  However, Refs.~[\onlinecite{Brandner2015Jan,Luo2018Aug}] considered idealized models of interactions, so much more work is necessary before experimental implementations could be imagined.

Thus, we argue for more experimental studies of potential barriers and point-contacts, since our model suggests that they could be the go-to solution for practical implementations of nanoscale heat engines and refrigerators.

\section{Acknowledgments}

C.C. thanks the QuanG program (HORIZON-MSCA-2021-COFUND-01) for the funding of her PhD.
R.W. acknowledges the support of the French National Research Agency (ANR)
through the project ``TQT'' (ANR-20-CE30-0028),
the project ``QuRes'' (ANR-21-CE47-0019), and 
the OECQ project (Contract DOS0226235/00) that is financed by the French state (BPI-France and France 2030) and Next Generation EU (via France Relance).


\appendix
\section{Codes and a trick for numerics}
\label{sec:numerics}

The python codes and data used to generate the figures in this article are available at Ref.~[\onlinecite{zenodo-codes}].

When performing the numerics it is worth noting that Eqs.~(\ref{eq:IL}-\ref{eq:eta-fri}) contain integrals which only converge at $\epsilon\to-\infty$ because the integrand contains a difference of two terms, $f_{\rm L}(\epsilon)$ and $f_{\rm R}(\epsilon)$, which each individually go to one as $\epsilon\to -\infty$.
Thus, any numerical algorithm that calculates the two terms in the integral separately will struggle to get a convergent answer.  There are various ways to cure this problem,
but an elegant one is to treat all states with energies below reservoir L's electrochemical potential as holes rather than electrons.
As we have chosen our scale of energy so that reservoir L's electrochemical potential, $\mu_{\rm L}=0$, we rewrite all electron states with negative $\epsilon$ as hole states with positive $\epsilon$, see for example sec.~4.5 of Ref.~[\onlinecite{Benenti2017Jun}].
This relies on the fact that Fermi functions take the form $f(x)=1/(1+\e^{x})$,
and so we can use $f(-x)= 1-f(x)$ to write the negative electron energies in terms of the positive hole energies.
Then 
\begin{eqnarray}
I_{\rm L}&=&\frac{1}{h} \sum_{\varsigma=\pm 1} \int_0^{\infty} \mathrm{d} \epsilon  \ \mathcal{T}_{\varsigma}(\epsilon)\left[f_{\rm L}^\varsigma(\epsilon)-f_{\rm R}^\varsigma(\epsilon)\right],
\label{eq:I-numerics}
\\
J_{\rm L}&=&\frac{1}{h} \sum_{\varsigma=\pm 1} \int_0^{\infty} \mathrm{d} \epsilon \ \epsilon\  \mathcal{T}_{\varsigma}(\epsilon)\left[f_{\rm L}^\varsigma(\epsilon)-f_{\rm R}^\varsigma(\epsilon)\right].
\label{eq:J-numerics}
\end{eqnarray}
Here, $\varsigma=+1$ corresponds to electrons (i.e., particles at an energy $\epsilon$ above the electrochemical potential of reservoir L), and $\varsigma=-1$ corresponds to holes (i.e.,  particles at an energy $\epsilon$ below the electrochemical potential of reservoir L). 
The function $f_i^\varsigma(\epsilon)$ corresponds to the Fermi function for a particle of type $\varsigma$ in reservoir $i$,
\begin{equation}
f_i^\varsigma(\epsilon)=\Big(1+\exp \big[\left(\epsilon-\varsigma \mu_i\right) /\left(k_{\mathrm{B}} T_i\right)\big]\Big)^{-1}.
\end{equation}
The function $\mathcal{T}_{\varsigma}(\epsilon)$ is the probability that a particle of type $\varsigma$ at energy $\epsilon$ transmits from one reservoir to the other.
This is found by taking the original transmission function
$\mathcal{T}(\epsilon)$ for electrons at energy $\epsilon$
above or below the electrochemical potential of reservoir L, as given in Eqs.~(\ref{eq:Tbarr},\ref{eq:Tdot}),
and cutting it at $\epsilon=0$ into two transmission functions.
In other words, if ${\cal T}(\epsilon)$ is the transmission for the full spread of energies from $\epsilon=-\infty$ to $\epsilon=\infty$, then 
\begin{eqnarray}
{\cal T}_{+1}(\epsilon) &\equiv&  {\cal T}(\epsilon)   \, ,
\\
{\cal T}_{-1}(\epsilon) &\equiv&  {\cal T}(-\epsilon)  \, ,
\end{eqnarray}
for all $\epsilon\geq 0$.
As before, the system generates power $P_{\text{gen}}=I_{\rm L}\mu_{\rm R}$, because we have defined $\mu_{\rm L}=0$. 

The approach that gives Eqs.~(\ref{eq:I-numerics},\ref{eq:J-numerics}) was originally developed to simplify the treatment of superconducting reservoirs.\cite{Benenti2017Jun} It allows one to model an electron at energy $\epsilon$ hitting the superconducting reservoir and being Andreev reflected as a hole with energy $\epsilon$.
Here we have no superconductor, and so no Andreev reflection, but we see that the equations in the form in Eqs.~(\ref{eq:I-numerics},\ref{eq:J-numerics}) have no terms in their integrand that causes an integral to diverge at the limits of integration, indeed all terms converge exponentially as $\epsilon \to \infty$. 
As a result, writing the equations in this form allowed us to use the standard Python packages for numerical integration without any risk of poor results due to a lack of convergence.

\section{Inverting $x=y e^{-y}$ for small $x$}
\label{app:inverting}

We want to invert $x=y e^{-y}$ to find $y$ as a function of $x$, when we are interested in the limit of small $x$.
This inverse is a special function called the Lambert W function, and it has been much studied.
However, here we only need one limit of it, which we can find without turning to books on special functions.
Plotting $y$ as a function of $x$, one sees graphically that it is a multi-valued function, with two values of $y$ at each value of $x$ up to $x=0.37$, and no values for larger $x$.
At small $x$, it has one solution at small $y$ (of the form $y=x +{\cal O}[x^2]$),
but it is the {\it other} solution that interests us;
we need to find the value of $y$ at small $x$ which corresponds to large $y$. 

To proceed, we actually consider the equality
\begin{eqnarray}
x=y^\alpha e^{-y},
\end{eqnarray}
and take $\alpha\to 1$ at the end.
Taking the logarithm of both side of this equality
gives
\begin{eqnarray}
-\ln[x]=y- \alpha \ln[y].
\label{eq:log-of-function-to-invert}
\end{eqnarray}
We assume that the solution we are looking for (for $y$ as a function of $x$) takes the form
\begin{eqnarray}
y(x)=f_0(x) + \alpha f_1(x) + \alpha^2 f_2(x) + \alpha^3 f_3(x) + \cdots \, .\ \ 
\end{eqnarray}
We substitute this into  Eq.~(\ref{eq:log-of-function-to-invert}) and expand the logarithm in powers of $\alpha$.  Then we write the equality for each power of $\alpha$.
The equalities between terms at zeroth, first, second and third order in $\alpha$, respectively give
\begin{eqnarray}
f_0(x) &=& \ln (1/x)\, ,\\
f_1(x) &=& \ln \big(f_0(x)\big) \, ,\\
f_2(x) &=& \ln f_1(x)\big/f_0(x)\, ,\\
f_3(x) &=& \ln f_2(x)\big/f_0(x) - f_1(x)\big/f_0(x))\, .
\end{eqnarray}
Then defining ${\cal L}(x)\equiv  \ln\big(\ln(1/x)\big)$ for compactness, we get
\begin{eqnarray}
y(x) &=& \ln (1/x) 
+ \alpha\bigg({\cal L}(x) 
+ \frac{\alpha}{\ln(1/x)}  {\cal L}(x)
\nonumber \\
& & \qquad \qquad \qquad
+ \left[\frac{\alpha}{\ln(1/x)}\right]^{\!2} {\cal L}(x)\Big[2-{\cal L}(x)\Big] 
\nonumber \\
& & \qquad\qquad \qquad \ \ 
+ {\cal O}\left[\left[\alpha\big/\ln(1/x)\right]^{\!3} \right] \bigg).\qquad
\end{eqnarray}
Taking the expansion up to second order in $\alpha$, and then setting $\alpha=1$, we get 
the approximation
\begin{eqnarray}
y(x) &\simeq& \ln (1/x)  
+ \left(1+ \frac{1}{\ln(1/x)}\right)  \ln\big(\ln(1/x)\big). \qquad
\label{eq:approx-inverse}
\end{eqnarray}
Plotting this approximation on the exact result shows that it is excellent at $x\to 0$; deviations from the exact result grow with $x$, but it works well for $x$ up to $0.2$ (only a few percent error at $x=0.2$). Eq.~(\ref{eq:approx-inverse}) is good enough for our purposes, but if one ever needs a better approximation then one can include the third order term in $\alpha$ before setting  $\alpha=1$, it then works well for $x$ as large as $0.3$ (only a few percent error at $x=0.3$). 


\bibliography{cite}

\begin{thebibliography}{61}%
\makeatletter
\providecommand \@ifxundefined [1]{%
 \@ifx{#1\undefined}
}%
\providecommand \@ifnum [1]{%
 \ifnum #1\expandafter \@firstoftwo
 \else \expandafter \@secondoftwo
 \fi
}%
\providecommand \@ifx [1]{%
 \ifx #1\expandafter \@firstoftwo
 \else \expandafter \@secondoftwo
 \fi
}%
\providecommand \natexlab [1]{#1}%
\providecommand \enquote  [1]{``#1''}%
\providecommand \bibnamefont  [1]{#1}%
\providecommand \bibfnamefont [1]{#1}%
\providecommand \citenamefont [1]{#1}%
\providecommand \href@noop [0]{\@secondoftwo}%
\providecommand \href [0]{\begingroup \@sanitize@url \@href}%
\providecommand \@href[1]{\@@startlink{#1}\@@href}%
\providecommand \@@href[1]{\endgroup#1\@@endlink}%
\providecommand \@sanitize@url [0]{\catcode `\\12\catcode `\$12\catcode
  `\&12\catcode `\#12\catcode `\^12\catcode `\_12\catcode `\%12\relax}%
\providecommand \@@startlink[1]{}%
\providecommand \@@endlink[0]{}%
\providecommand \url  [0]{\begingroup\@sanitize@url \@url }%
\providecommand \@url [1]{\endgroup\@href {#1}{\urlprefix }}%
\providecommand \urlprefix  [0]{URL }%
\providecommand \Eprint [0]{\href }%
\providecommand \doibase [0]{http://dx.doi.org/}%
\providecommand \selectlanguage [0]{\@gobble}%
\providecommand \bibinfo  [0]{\@secondoftwo}%
\providecommand \bibfield  [0]{\@secondoftwo}%
\providecommand \translation [1]{[#1]}%
\providecommand \BibitemOpen [0]{}%
\providecommand \bibitemStop [0]{}%
\providecommand \bibitemNoStop [0]{.\EOS\space}%
\providecommand \EOS [0]{\spacefactor3000\relax}%
\providecommand \BibitemShut  [1]{\csname bibitem#1\endcsname}%
\let\auto@bib@innerbib\@empty
\bibitem [{\citenamefont {Callen}(1985)}]{Callen1985Sep}%
  \BibitemOpen
  \bibfield  {author} {\bibinfo {author} {\bibfnamefont {Herbert~B.}\
  \bibnamefont {Callen}},\ }\href@noop {} {\emph {\bibinfo {title}
  {{Thermodynamics and an Introduction to Thermostatistics}}}}\ (\bibinfo
  {publisher} {Wiley},\ \bibinfo {year} {1985})\BibitemShut {NoStop}%
\bibitem [{\citenamefont {Mahan}\ and\ \citenamefont
  {Sofo}(1996)}]{Mahan1996Jul}%
  \BibitemOpen
  \bibfield  {author} {\bibinfo {author} {\bibfnamefont {G.~D.}\ \bibnamefont
  {Mahan}}\ and\ \bibinfo {author} {\bibfnamefont {J.~O.}\ \bibnamefont
  {Sofo}},\ }\bibfield  {title} {\enquote {\bibinfo {title} {{The best
  thermoelectric.}}}\ }\href {\doibase 10.1073/pnas.93.15.7436} {\bibfield
  {journal} {\bibinfo  {journal} {Proc. Natl. Acad. Sci. U.S.A.}\ }\textbf
  {\bibinfo {volume} {93}},\ \bibinfo {pages} {7436--7439} (\bibinfo {year}
  {1996})}\BibitemShut {NoStop}%
\bibitem [{\citenamefont {Humphrey}\ and\ \citenamefont
  {Linke}(2005)}]{Humphrey2005Mar}%
  \BibitemOpen
  \bibfield  {author} {\bibinfo {author} {\bibfnamefont {T.~E.}\ \bibnamefont
  {Humphrey}}\ and\ \bibinfo {author} {\bibfnamefont {H.}~\bibnamefont
  {Linke}},\ }\bibfield  {title} {\enquote {\bibinfo {title} {Reversible
  thermoelectric nanomaterials},}\ }\href {\doibase
  10.1103/PhysRevLett.94.096601} {\bibfield  {journal} {\bibinfo  {journal}
  {Phys. Rev. Lett.}\ }\textbf {\bibinfo {volume} {94}},\ \bibinfo {pages}
  {096601} (\bibinfo {year} {2005})}\BibitemShut {NoStop}%
\bibitem [{\citenamefont {Whitney}(2014)}]{Whitney2014Apr}%
  \BibitemOpen
  \bibfield  {author} {\bibinfo {author} {\bibfnamefont {Robert~S.}\
  \bibnamefont {Whitney}},\ }\bibfield  {title} {\enquote {\bibinfo {title}
  {{Most Efficient Quantum Thermoelectric at Finite Power Output}},}\ }\href
  {\doibase 10.1103/PhysRevLett.112.130601} {\bibfield  {journal} {\bibinfo
  {journal} {Phys. Rev. Lett.}\ }\textbf {\bibinfo {volume} {112}},\ \bibinfo
  {pages} {130601} (\bibinfo {year} {2014})}\BibitemShut {NoStop}%
\bibitem [{\citenamefont {Whitney}(2015)}]{Whitney2015Mar}%
  \BibitemOpen
  \bibfield  {author} {\bibinfo {author} {\bibfnamefont {Robert~S.}\
  \bibnamefont {Whitney}},\ }\bibfield  {title} {\enquote {\bibinfo {title}
  {{Finding the quantum thermoelectric with maximal efficiency and minimal
  entropy production at given power output}},}\ }\href {\doibase
  10.1103/PhysRevB.91.115425} {\bibfield  {journal} {\bibinfo  {journal} {Phys.
  Rev. B}\ }\textbf {\bibinfo {volume} {91}},\ \bibinfo {pages} {115425}
  (\bibinfo {year} {2015})}\BibitemShut {NoStop}%
\bibitem [{\citenamefont {Whitney}(2016)}]{Whitney2016May}%
  \BibitemOpen
  \bibfield  {author} {\bibinfo {author} {\bibfnamefont {Robert~S.}\
  \bibnamefont {Whitney}},\ }\bibfield  {title} {\enquote {\bibinfo {title}
  {Quantum coherent three-terminal thermoelectrics: Maximum efficiency at given
  power output},}\ }\href {\doibase 10.3390/e18060208} {\bibfield  {journal}
  {\bibinfo  {journal} {Entropy}\ }\textbf {\bibinfo {volume} {18}},\ \bibinfo
  {pages} {208} (\bibinfo {year} {2016})}\BibitemShut {NoStop}%
\bibitem [{\citenamefont {Ding}\ \emph {et~al.}(2023)\citenamefont {Ding},
  \citenamefont {Chen}, \citenamefont {Xu},\ and\ \citenamefont
  {Duan}}]{Ding2023Oct}%
  \BibitemOpen
  \bibfield  {author} {\bibinfo {author} {\bibfnamefont {Sifan}\ \bibnamefont
  {Ding}}, \bibinfo {author} {\bibfnamefont {Xiaobin}\ \bibnamefont {Chen}},
  \bibinfo {author} {\bibfnamefont {Yong}\ \bibnamefont {Xu}}, \ and\ \bibinfo
  {author} {\bibfnamefont {Wenhui}\ \bibnamefont {Duan}},\ }\bibfield  {title}
  {\enquote {\bibinfo {title} {{The best thermoelectrics revisited in the
  quantum limit}},}\ }\href {\doibase 10.1038/s41524-023-01141-1} {\bibfield
  {journal} {\bibinfo  {journal} {npj Comput. Mater.}\ }\textbf {\bibinfo
  {volume} {9}},\ \bibinfo {pages} {189} (\bibinfo {year} {2023})}\BibitemShut
  {NoStop}%
\bibitem [{\citenamefont {Maassen}(2021)}]{Maassen2021Nov}%
  \BibitemOpen
  \bibfield  {author} {\bibinfo {author} {\bibfnamefont {Jesse}\ \bibnamefont
  {Maassen}},\ }\bibfield  {title} {\enquote {\bibinfo {title} {{Limits of
  thermoelectric performance with a bounded transport distribution}},}\ }\href
  {\doibase 10.1103/PhysRevB.104.184301} {\bibfield  {journal} {\bibinfo
  {journal} {Phys. Rev. B}\ }\textbf {\bibinfo {volume} {104}},\ \bibinfo
  {pages} {184301} (\bibinfo {year} {2021})}\BibitemShut {NoStop}%
\bibitem [{Note1()}]{Note1}%
  \BibitemOpen
  \bibinfo {note} {Scattering theory holds for non-interacting electrons, or
  when inter-particle interactions are treated at a mean-field (Hartree) level.
  Ref.~[\protect \rev@citealpnum {Maassen2021Nov}] uses the linear Boltzmann
  transport equation within the relaxation time approximation. The question of
  a bound on efficiency at finite power remains open in other systems, with
  this bound exceeded in simulations of some systems with strong
  relaxation\cite {Brandner2015Jan} or inter-particle interactions.\cite
  {Luo2018Aug}}\BibitemShut {NoStop}%
\bibitem [{\citenamefont {Maiti}\ and\ \citenamefont
  {Nitzan}(2013)}]{maiti2013mobility}%
  \BibitemOpen
  \bibfield  {author} {\bibinfo {author} {\bibfnamefont {Santanu~K}\
  \bibnamefont {Maiti}}\ and\ \bibinfo {author} {\bibfnamefont {Abraham}\
  \bibnamefont {Nitzan}},\ }\bibfield  {title} {\enquote {\bibinfo {title}
  {{Mobility edge phenomenon in a Hubbard chain: A mean field study}},}\ }\href
  {\doibase 10.1016/j.physleta.2013.03.013} {\bibfield  {journal} {\bibinfo
  {journal} {Phys. Lett. A}\ }\textbf {\bibinfo {volume} {377}},\ \bibinfo
  {pages} {1205} (\bibinfo {year} {2013})}\BibitemShut {NoStop}%
\bibitem [{\citenamefont {Yamamoto}\ \emph {et~al.}(2017)\citenamefont
  {Yamamoto}, \citenamefont {Aharony}, \citenamefont {Entin-Wohlman},\ and\
  \citenamefont {Hatano}}]{yamamoto2017thermoelectricity}%
  \BibitemOpen
  \bibfield  {author} {\bibinfo {author} {\bibfnamefont {Kaoru}\ \bibnamefont
  {Yamamoto}}, \bibinfo {author} {\bibfnamefont {Amnon}\ \bibnamefont
  {Aharony}}, \bibinfo {author} {\bibfnamefont {Ora}\ \bibnamefont
  {Entin-Wohlman}}, \ and\ \bibinfo {author} {\bibfnamefont {Naomichi}\
  \bibnamefont {Hatano}},\ }\bibfield  {title} {\enquote {\bibinfo {title}
  {Thermoelectricity near {A}nderson localization transitions},}\ }\href
  {\doibase 10.1103/PhysRevB.96.155201} {\bibfield  {journal} {\bibinfo
  {journal} {Phys. Rev. B}\ }\textbf {\bibinfo {volume} {96}},\ \bibinfo
  {pages} {155201} (\bibinfo {year} {2017})}\BibitemShut {NoStop}%
\bibitem [{\citenamefont {Samuelsson}\ \emph {et~al.}(2017)\citenamefont
  {Samuelsson}, \citenamefont {Kheradsoud},\ and\ \citenamefont
  {Sothmann}}]{samuelsson2017optimal}%
  \BibitemOpen
  \bibfield  {author} {\bibinfo {author} {\bibfnamefont {Peter}\ \bibnamefont
  {Samuelsson}}, \bibinfo {author} {\bibfnamefont {Sara}\ \bibnamefont
  {Kheradsoud}}, \ and\ \bibinfo {author} {\bibfnamefont {Bj{\"o}rn}\
  \bibnamefont {Sothmann}},\ }\bibfield  {title} {\enquote {\bibinfo {title}
  {Optimal quantum interference thermoelectric heat engine with edge states},}\
  }\href {\doibase 10.1103/PhysRevLett.118.256801} {\bibfield  {journal}
  {\bibinfo  {journal} {Phys. Rev. Lett.}\ }\textbf {\bibinfo {volume} {118}},\
  \bibinfo {pages} {256801} (\bibinfo {year} {2017})}\BibitemShut {NoStop}%
\bibitem [{\citenamefont {Haack}\ and\ \citenamefont
  {Giazotto}(2019)}]{haack2019efficient}%
  \BibitemOpen
  \bibfield  {author} {\bibinfo {author} {\bibfnamefont {G{\'e}raldine}\
  \bibnamefont {Haack}}\ and\ \bibinfo {author} {\bibfnamefont {Francesco}\
  \bibnamefont {Giazotto}},\ }\bibfield  {title} {\enquote {\bibinfo {title}
  {{Efficient and tunable Aharonov-Bohm quantum heat engine}},}\ }\href
  {\doibase 10.1103/PhysRevB.100.235442} {\bibfield  {journal} {\bibinfo
  {journal} {Phys. Rev. B}\ }\textbf {\bibinfo {volume} {100}},\ \bibinfo
  {pages} {235442} (\bibinfo {year} {2019})}\BibitemShut {NoStop}%
\bibitem [{\citenamefont {Chiaracane}\ \emph {et~al.}(2020)\citenamefont
  {Chiaracane}, \citenamefont {Mitchison}, \citenamefont {Purkayastha},
  \citenamefont {Haack},\ and\ \citenamefont
  {Goold}}]{chiaracane2020quasiperiodic}%
  \BibitemOpen
  \bibfield  {author} {\bibinfo {author} {\bibfnamefont {Cecilia}\ \bibnamefont
  {Chiaracane}}, \bibinfo {author} {\bibfnamefont {Mark~T}\ \bibnamefont
  {Mitchison}}, \bibinfo {author} {\bibfnamefont {Archak}\ \bibnamefont
  {Purkayastha}}, \bibinfo {author} {\bibfnamefont {G{\'e}raldine}\
  \bibnamefont {Haack}}, \ and\ \bibinfo {author} {\bibfnamefont {John}\
  \bibnamefont {Goold}},\ }\bibfield  {title} {\enquote {\bibinfo {title}
  {Quasiperiodic quantum heat engines with a mobility edge},}\ }\href {\doibase
  10.1103/PhysRevResearch.2.013093} {\bibfield  {journal} {\bibinfo  {journal}
  {Phys. Rev. Research}\ }\textbf {\bibinfo {volume} {2}},\ \bibinfo {pages}
  {013093} (\bibinfo {year} {2020})}\BibitemShut {NoStop}%
\bibitem [{\citenamefont {Haack}\ and\ \citenamefont
  {Giazotto}(2021)}]{haack2021nonlinear}%
  \BibitemOpen
  \bibfield  {author} {\bibinfo {author} {\bibfnamefont {G{\'e}raldine}\
  \bibnamefont {Haack}}\ and\ \bibinfo {author} {\bibfnamefont {Francesco}\
  \bibnamefont {Giazotto}},\ }\bibfield  {title} {\enquote {\bibinfo {title}
  {Nonlinear regime for enhanced performance of an {Aharonov--Bohm} heat
  engine},}\ }\href {\doibase 10.1116/5.0064936} {\bibfield  {journal}
  {\bibinfo  {journal} {AVS Quantum Science}\ }\textbf {\bibinfo {volume}
  {3}},\ \bibinfo {pages} {046801} (\bibinfo {year} {2021})}\BibitemShut
  {NoStop}%
\bibitem [{\citenamefont {Behera}\ \emph {et~al.}(2023)\citenamefont {Behera},
  \citenamefont {Bedkihal}, \citenamefont {Agarwalla},\ and\ \citenamefont
  {Bandyopadhyay}}]{behera2023quantum}%
  \BibitemOpen
  \bibfield  {author} {\bibinfo {author} {\bibfnamefont {Jayasmita}\
  \bibnamefont {Behera}}, \bibinfo {author} {\bibfnamefont {Salil}\
  \bibnamefont {Bedkihal}}, \bibinfo {author} {\bibfnamefont {Bijay~Kumar}\
  \bibnamefont {Agarwalla}}, \ and\ \bibinfo {author} {\bibfnamefont {Malay}\
  \bibnamefont {Bandyopadhyay}},\ }\bibfield  {title} {\enquote {\bibinfo
  {title} {Quantum coherent control of nonlinear thermoelectric transport in a
  triple-dot aharonov-bohm heat engine},}\ }\href {\doibase
  10.1103/PhysRevB.108.165419} {\bibfield  {journal} {\bibinfo  {journal}
  {Phys. Rev. B}\ }\textbf {\bibinfo {volume} {108}},\ \bibinfo {pages}
  {165419} (\bibinfo {year} {2023})}\BibitemShut {NoStop}%
\bibitem [{\citenamefont {Brandner}\ and\ \citenamefont
  {Saito}(2025)}]{brandner2025thermodynamic}%
  \BibitemOpen
  \bibfield  {author} {\bibinfo {author} {\bibfnamefont {Kay}\ \bibnamefont
  {Brandner}}\ and\ \bibinfo {author} {\bibfnamefont {Keiji}\ \bibnamefont
  {Saito}},\ }\bibfield  {title} {\enquote {\bibinfo {title} {Thermodynamic
  uncertainty relations for coherent transport},}\ }\href@noop {} {\bibfield
  {journal} {\bibinfo  {journal} {Preprint - arXiv:2502.07917}\ } (\bibinfo
  {year} {2025})}\BibitemShut {NoStop}%
\bibitem [{\citenamefont {Chen}\ \emph {et~al.}(2018)\citenamefont {Chen},
  \citenamefont {Burke}, \citenamefont {Svilans}, \citenamefont {Linke},\ and\
  \citenamefont {Thelander}}]{Chen2018-experiment}%
  \BibitemOpen
  \bibfield  {author} {\bibinfo {author} {\bibfnamefont {I-Ju}\ \bibnamefont
  {Chen}}, \bibinfo {author} {\bibfnamefont {Adam}\ \bibnamefont {Burke}},
  \bibinfo {author} {\bibfnamefont {Artis}\ \bibnamefont {Svilans}}, \bibinfo
  {author} {\bibfnamefont {Heiner}\ \bibnamefont {Linke}}, \ and\ \bibinfo
  {author} {\bibfnamefont {Claes}\ \bibnamefont {Thelander}},\ }\bibfield
  {title} {\enquote {\bibinfo {title} {Thermoelectric power factor limit of a
  1d nanowire},}\ }\href {\doibase 10.1103/PhysRevLett.120.177703} {\bibfield
  {journal} {\bibinfo  {journal} {Phys. Rev. Lett.}\ }\textbf {\bibinfo
  {volume} {120}},\ \bibinfo {pages} {177703} (\bibinfo {year}
  {2018})}\BibitemShut {NoStop}%
\bibitem [{\citenamefont {Reihani}\ \emph {et~al.}(2024)\citenamefont
  {Reihani}, \citenamefont {Li}, \citenamefont {Guan}, \citenamefont {Luan},
  \citenamefont {Yan}, \citenamefont {Xue}, \citenamefont {Meyhofer},
  \citenamefont {Reddy},\ and\ \citenamefont {Ram}}]{reihani2024cooling}%
  \BibitemOpen
  \bibfield  {author} {\bibinfo {author} {\bibfnamefont {Amin}\ \bibnamefont
  {Reihani}}, \bibinfo {author} {\bibfnamefont {Zheng}\ \bibnamefont {Li}},
  \bibinfo {author} {\bibfnamefont {Jian}\ \bibnamefont {Guan}}, \bibinfo
  {author} {\bibfnamefont {Yuxuan}\ \bibnamefont {Luan}}, \bibinfo {author}
  {\bibfnamefont {Shen}\ \bibnamefont {Yan}}, \bibinfo {author} {\bibfnamefont
  {Jin}\ \bibnamefont {Xue}}, \bibinfo {author} {\bibfnamefont {Edgar}\
  \bibnamefont {Meyhofer}}, \bibinfo {author} {\bibfnamefont {Pramod}\
  \bibnamefont {Reddy}}, \ and\ \bibinfo {author} {\bibfnamefont {Rajeev~J}\
  \bibnamefont {Ram}},\ }\bibfield  {title} {\enquote {\bibinfo {title}
  {Cooling of semiconductor devices via quantum tunneling},}\ }\href@noop {}
  {\bibfield  {journal} {\bibinfo  {journal} {Physical review letters}\
  }\textbf {\bibinfo {volume} {133}},\ \bibinfo {pages} {266301} (\bibinfo
  {year} {2024})}\BibitemShut {NoStop}%
\bibitem [{\citenamefont {Mykk{\ifmmode\ddot{a}\else\"{a}\fi}nen}\ \emph
  {et~al.}(2020)\citenamefont {Mykk{\ifmmode\ddot{a}\else\"{a}\fi}nen},
  \citenamefont {Lehtinen}, \citenamefont
  {Gr{\ifmmode\ddot{o}\else\"{o}\fi}nberg}, \citenamefont {Shchepetov},
  \citenamefont {Timofeev}, \citenamefont {Gunnarsson}, \citenamefont
  {Kemppinen}, \citenamefont {Manninen},\ and\ \citenamefont
  {Prunnila}}]{Mykkanen2020Apr}%
  \BibitemOpen
  \bibfield  {author} {\bibinfo {author} {\bibfnamefont {Emma}\ \bibnamefont
  {Mykk{\ifmmode\ddot{a}\else\"{a}\fi}nen}}, \bibinfo {author} {\bibfnamefont
  {Janne~S.}\ \bibnamefont {Lehtinen}}, \bibinfo {author} {\bibfnamefont
  {Leif}\ \bibnamefont {Gr{\ifmmode\ddot{o}\else\"{o}\fi}nberg}}, \bibinfo
  {author} {\bibfnamefont {Andrey}\ \bibnamefont {Shchepetov}}, \bibinfo
  {author} {\bibfnamefont {Andrey~V.}\ \bibnamefont {Timofeev}}, \bibinfo
  {author} {\bibfnamefont {David}\ \bibnamefont {Gunnarsson}}, \bibinfo
  {author} {\bibfnamefont {Antti}\ \bibnamefont {Kemppinen}}, \bibinfo {author}
  {\bibfnamefont {Antti~J.}\ \bibnamefont {Manninen}}, \ and\ \bibinfo {author}
  {\bibfnamefont {Mika}\ \bibnamefont {Prunnila}},\ }\bibfield  {title}
  {\enquote {\bibinfo {title} {Thermionic junction devices utilizing phonon
  blocking},}\ }\href {\doibase 10.1126/sciadv.aax9191} {\bibfield  {journal}
  {\bibinfo  {journal} {Sci. Adv.}\ }\textbf {\bibinfo {volume} {6}},\ \bibinfo
  {pages} {eaax9191} (\bibinfo {year} {2020})}\BibitemShut {NoStop}%
\bibitem [{\citenamefont {Shakouri}\ and\ \citenamefont
  {Bowers}(1997)}]{Shakouri1997Sep}%
  \BibitemOpen
  \bibfield  {author} {\bibinfo {author} {\bibfnamefont {Ali}\ \bibnamefont
  {Shakouri}}\ and\ \bibinfo {author} {\bibfnamefont {John~E.}\ \bibnamefont
  {Bowers}},\ }\bibfield  {title} {\enquote {\bibinfo {title} {{Heterostructure
  integrated thermionic coolers}},}\ }\href {\doibase 10.1063/1.119861}
  {\bibfield  {journal} {\bibinfo  {journal} {Appl. Phys. Lett.}\ }\textbf
  {\bibinfo {volume} {71}},\ \bibinfo {pages} {1234} (\bibinfo {year}
  {1997})}\BibitemShut {NoStop}%
\bibitem [{\citenamefont {Shakouri}\ \emph {et~al.}(1998)\citenamefont
  {Shakouri}, \citenamefont {LaBounty}, \citenamefont {Abraham}, \citenamefont
  {Piprek},\ and\ \citenamefont {Bowers}}]{Shakouri1998Jan}%
  \BibitemOpen
  \bibfield  {author} {\bibinfo {author} {\bibfnamefont {Ali}\ \bibnamefont
  {Shakouri}}, \bibinfo {author} {\bibfnamefont {Chris}\ \bibnamefont
  {LaBounty}}, \bibinfo {author} {\bibfnamefont {Patrick}\ \bibnamefont
  {Abraham}}, \bibinfo {author} {\bibfnamefont {Joachim}\ \bibnamefont
  {Piprek}}, \ and\ \bibinfo {author} {\bibfnamefont {John~E.}\ \bibnamefont
  {Bowers}},\ }\bibfield  {title} {\enquote {\bibinfo {title} {Enhanced
  thermionic emission cooling in high barrier superlattice heterostructures},}\
  }\href {\doibase 10.1557/PROC-545-449} {\bibfield  {journal} {\bibinfo
  {journal} {MRS Online Proceedings Library (OPL)}\ }\textbf {\bibinfo {volume}
  {545}},\ \bibinfo {pages} {449} (\bibinfo {year} {1998})}\BibitemShut
  {NoStop}%
\bibitem [{\citenamefont {Molenkamp}\ \emph {et~al.}(1990)\citenamefont
  {Molenkamp}, \citenamefont {van Houten}, \citenamefont {Beenakker},
  \citenamefont {Eppenga},\ and\ \citenamefont {Foxon}}]{Molenkamp1990Aug}%
  \BibitemOpen
  \bibfield  {author} {\bibinfo {author} {\bibfnamefont {L.~W.}\ \bibnamefont
  {Molenkamp}}, \bibinfo {author} {\bibfnamefont {H.}~\bibnamefont {van
  Houten}}, \bibinfo {author} {\bibfnamefont {C.~W.~J.}\ \bibnamefont
  {Beenakker}}, \bibinfo {author} {\bibfnamefont {R.}~\bibnamefont {Eppenga}},
  \ and\ \bibinfo {author} {\bibfnamefont {C.~T.}\ \bibnamefont {Foxon}},\
  }\bibfield  {title} {\enquote {\bibinfo {title} {{Quantum oscillations in the
  transverse voltage of a channel in the nonlinear transport regime}},}\ }\href
  {\doibase 10.1103/PhysRevLett.65.1052} {\bibfield  {journal} {\bibinfo
  {journal} {Phys. Rev. Lett.}\ }\textbf {\bibinfo {volume} {65}},\ \bibinfo
  {pages} {1052} (\bibinfo {year} {1990})}\BibitemShut {NoStop}%
\bibitem [{\citenamefont {Dzurak}\ \emph {et~al.}(1993)\citenamefont {Dzurak},
  \citenamefont {Smith}, \citenamefont {Martin-Moreno}, \citenamefont {Pepper},
  \citenamefont {Ritchie}, \citenamefont {Jones},\ and\ \citenamefont
  {Hasko}}]{Dzurak1993Oct}%
  \BibitemOpen
  \bibfield  {author} {\bibinfo {author} {\bibfnamefont {A.~S.}\ \bibnamefont
  {Dzurak}}, \bibinfo {author} {\bibfnamefont {C.~G.}\ \bibnamefont {Smith}},
  \bibinfo {author} {\bibfnamefont {L.}~\bibnamefont {Martin-Moreno}}, \bibinfo
  {author} {\bibfnamefont {M.}~\bibnamefont {Pepper}}, \bibinfo {author}
  {\bibfnamefont {D.~A.}\ \bibnamefont {Ritchie}}, \bibinfo {author}
  {\bibfnamefont {G.~A.~C.}\ \bibnamefont {Jones}}, \ and\ \bibinfo {author}
  {\bibfnamefont {D.~G.}\ \bibnamefont {Hasko}},\ }\bibfield  {title} {\enquote
  {\bibinfo {title} {{Thermopower of a one-dimensional ballistic constriction
  in the non-linear regime}},}\ }\href {\doibase 10.1088/0953-8984/5/43/017}
  {\bibfield  {journal} {\bibinfo  {journal} {J. Phys.: Condens. Matter}\
  }\textbf {\bibinfo {volume} {5}},\ \bibinfo {pages} {8055} (\bibinfo {year}
  {1993})}\BibitemShut {NoStop}%
\bibitem [{\citenamefont {Brantut}\ \emph {et~al.}(2013)\citenamefont
  {Brantut}, \citenamefont {Grenier}, \citenamefont {Meineke}, \citenamefont
  {Stadler}, \citenamefont {Krinner}, \citenamefont {Kollath}, \citenamefont
  {Esslinger},\ and\ \citenamefont {Georges}}]{Brantut2013Nov}%
  \BibitemOpen
  \bibfield  {author} {\bibinfo {author} {\bibfnamefont {Jean-Philippe}\
  \bibnamefont {Brantut}}, \bibinfo {author} {\bibfnamefont {Charles}\
  \bibnamefont {Grenier}}, \bibinfo {author} {\bibfnamefont {Jakob}\
  \bibnamefont {Meineke}}, \bibinfo {author} {\bibfnamefont {David}\
  \bibnamefont {Stadler}}, \bibinfo {author} {\bibfnamefont {Sebastian}\
  \bibnamefont {Krinner}}, \bibinfo {author} {\bibfnamefont {Corinna}\
  \bibnamefont {Kollath}}, \bibinfo {author} {\bibfnamefont {Tilman}\
  \bibnamefont {Esslinger}}, \ and\ \bibinfo {author} {\bibfnamefont {Antoine}\
  \bibnamefont {Georges}},\ }\bibfield  {title} {\enquote {\bibinfo {title} {A
  thermoelectric heat engine with ultracold atoms},}\ }\href {\doibase
  10.1126/science.1242308} {\bibfield  {journal} {\bibinfo  {journal}
  {Science}\ }\textbf {\bibinfo {volume} {342}},\ \bibinfo {pages} {713}
  (\bibinfo {year} {2013})}\BibitemShut {NoStop}%
\bibitem [{foo()}]{footnote-atoms}%
  \BibitemOpen
  \href@noop {} {}\bibinfo {note} {{Such physics should be visible in atom trap
  experiments with a constriction created with a laser, but the constriction
  must be a bit narrower than that in
  Ref.~[\onlinecite{Brantut2013Nov}].}}\BibitemShut {Stop}%
\bibitem [{\citenamefont {Prance}\ \emph {et~al.}(2009)\citenamefont {Prance},
  \citenamefont {Smith}, \citenamefont {Griffiths}, \citenamefont {Chorley},
  \citenamefont {Anderson}, \citenamefont {Jones}, \citenamefont {Farrer},\
  and\ \citenamefont {Ritchie}}]{prance2009electronic}%
  \BibitemOpen
  \bibfield  {author} {\bibinfo {author} {\bibfnamefont {J.R.}\ \bibnamefont
  {Prance}}, \bibinfo {author} {\bibfnamefont {C.G.}\ \bibnamefont {Smith}},
  \bibinfo {author} {\bibfnamefont {J.P.}\ \bibnamefont {Griffiths}}, \bibinfo
  {author} {\bibfnamefont {S.J.}\ \bibnamefont {Chorley}}, \bibinfo {author}
  {\bibfnamefont {D.}~\bibnamefont {Anderson}}, \bibinfo {author}
  {\bibfnamefont {G.A.C.}\ \bibnamefont {Jones}}, \bibinfo {author}
  {\bibfnamefont {I.}~\bibnamefont {Farrer}}, \ and\ \bibinfo {author}
  {\bibfnamefont {D.A.}\ \bibnamefont {Ritchie}},\ }\bibfield  {title}
  {\enquote {\bibinfo {title} {Electronic refrigeration of a two-dimensional
  electron gas},}\ }\href {\doibase 10.1103/PhysRevLett.102.146602} {\bibfield
  {journal} {\bibinfo  {journal} {Phys. Rev. Lett.}\ }\textbf {\bibinfo
  {volume} {102}},\ \bibinfo {pages} {146602} (\bibinfo {year}
  {2009})}\BibitemShut {NoStop}%
\bibitem [{\citenamefont {Fahlvik~Svensson}\ \emph {et~al.}(2012)\citenamefont
  {Fahlvik~Svensson}, \citenamefont {Persson}, \citenamefont {Hoffmann},
  \citenamefont {Nakpathomkun}, \citenamefont {Nilsson}, \citenamefont {Xu},
  \citenamefont {Samuelson},\ and\ \citenamefont {Linke}}]{Fahlvik2012Mar}%
  \BibitemOpen
  \bibfield  {author} {\bibinfo {author} {\bibfnamefont {S.}~\bibnamefont
  {Fahlvik~Svensson}}, \bibinfo {author} {\bibfnamefont {A.~I.}\ \bibnamefont
  {Persson}}, \bibinfo {author} {\bibfnamefont {E.~A.}\ \bibnamefont
  {Hoffmann}}, \bibinfo {author} {\bibfnamefont {N.}~\bibnamefont
  {Nakpathomkun}}, \bibinfo {author} {\bibfnamefont {H.~A.}\ \bibnamefont
  {Nilsson}}, \bibinfo {author} {\bibfnamefont {H.~Q.}\ \bibnamefont {Xu}},
  \bibinfo {author} {\bibfnamefont {L.}~\bibnamefont {Samuelson}}, \ and\
  \bibinfo {author} {\bibfnamefont {H.}~\bibnamefont {Linke}},\ }\bibfield
  {title} {\enquote {\bibinfo {title} {{Lineshape of the thermopower of quantum
  dots}},}\ }\href {\doibase 10.1088/1367-2630/14/3/033041} {\bibfield
  {journal} {\bibinfo  {journal} {New J. Phys.}\ }\textbf {\bibinfo {volume}
  {14}},\ \bibinfo {pages} {033041} (\bibinfo {year} {2012})}\BibitemShut
  {NoStop}%
\bibitem [{\citenamefont {Josefsson}\ \emph {et~al.}(2018)\citenamefont
  {Josefsson}, \citenamefont {Svilans}, \citenamefont {Burke}, \citenamefont
  {Hoffmann}, \citenamefont {Fahlvik}, \citenamefont {Thelander}, \citenamefont
  {Leijnse},\ and\ \citenamefont {Linke}}]{Josefsson2018Oct}%
  \BibitemOpen
  \bibfield  {author} {\bibinfo {author} {\bibfnamefont {Martin}\ \bibnamefont
  {Josefsson}}, \bibinfo {author} {\bibfnamefont {Artis}\ \bibnamefont
  {Svilans}}, \bibinfo {author} {\bibfnamefont {Adam~M.}\ \bibnamefont
  {Burke}}, \bibinfo {author} {\bibfnamefont {Eric~A.}\ \bibnamefont
  {Hoffmann}}, \bibinfo {author} {\bibfnamefont {Sofia}\ \bibnamefont
  {Fahlvik}}, \bibinfo {author} {\bibfnamefont {Claes}\ \bibnamefont
  {Thelander}}, \bibinfo {author} {\bibfnamefont {Martin}\ \bibnamefont
  {Leijnse}}, \ and\ \bibinfo {author} {\bibfnamefont {Heiner}\ \bibnamefont
  {Linke}},\ }\bibfield  {title} {\enquote {\bibinfo {title} {{A quantum-dot
  heat engine operating close to the thermodynamic efficiency limits}},}\
  }\href {\doibase 10.1038/s41565-018-0200-5} {\bibfield  {journal} {\bibinfo
  {journal} {Nat. Nanotechnol.}\ }\textbf {\bibinfo {volume} {13}},\ \bibinfo
  {pages} {920--924} (\bibinfo {year} {2018})}\BibitemShut {NoStop}%
\bibitem [{\citenamefont {Reddy}\ \emph {et~al.}(2007)\citenamefont {Reddy},
  \citenamefont {Jang}, \citenamefont {Segalman},\ and\ \citenamefont
  {Majumdar}}]{Reddy2007Mar}%
  \BibitemOpen
  \bibfield  {author} {\bibinfo {author} {\bibfnamefont {Pramod}\ \bibnamefont
  {Reddy}}, \bibinfo {author} {\bibfnamefont {Sung-Yeon}\ \bibnamefont {Jang}},
  \bibinfo {author} {\bibfnamefont {Rachel~A.}\ \bibnamefont {Segalman}}, \
  and\ \bibinfo {author} {\bibfnamefont {Arun}\ \bibnamefont {Majumdar}},\
  }\bibfield  {title} {\enquote {\bibinfo {title} {Thermoelectricity in
  molecular junctions},}\ }\href {\doibase 10.1126/science.1137149} {\bibfield
  {journal} {\bibinfo  {journal} {Science}\ }\textbf {\bibinfo {volume}
  {315}},\ \bibinfo {pages} {1568} (\bibinfo {year} {2007})}\BibitemShut
  {NoStop}%
\bibitem [{\citenamefont {Cohen~Jungerman}\ \emph {et~al.}(2025)\citenamefont
  {Cohen~Jungerman}, \citenamefont {Shmueli}, \citenamefont {Shekhter},\ and\
  \citenamefont {Selzer}}]{cohen2025unusually}%
  \BibitemOpen
  \bibfield  {author} {\bibinfo {author} {\bibfnamefont {Mor}\ \bibnamefont
  {Cohen~Jungerman}}, \bibinfo {author} {\bibfnamefont {Shachar}\ \bibnamefont
  {Shmueli}}, \bibinfo {author} {\bibfnamefont {Pini}\ \bibnamefont
  {Shekhter}}, \ and\ \bibinfo {author} {\bibfnamefont {Yoram}\ \bibnamefont
  {Selzer}},\ }\bibfield  {title} {\enquote {\bibinfo {title} {Unusually high
  thermopower in molecular junctions from molecularly induced quantized states
  in their semimetal leads},}\ }\href {\doibase 10.1021/acs.nanolett.4c05852}
  {\bibfield  {journal} {\bibinfo  {journal} {Nano Lett.}\ }\textbf {\bibinfo
  {volume} {25}},\ \bibinfo {pages} {2756} (\bibinfo {year}
  {2025})}\BibitemShut {NoStop}%
\bibitem [{\citenamefont {Edwards}\ \emph {et~al.}(1993)\citenamefont
  {Edwards}, \citenamefont {Niu},\ and\ \citenamefont
  {De~Lozanne}}]{edwards1993quantum}%
  \BibitemOpen
  \bibfield  {author} {\bibinfo {author} {\bibfnamefont {H.L.}\ \bibnamefont
  {Edwards}}, \bibinfo {author} {\bibfnamefont {Q.D.}\ \bibnamefont {Niu}}, \
  and\ \bibinfo {author} {\bibfnamefont {A.L.}\ \bibnamefont {De~Lozanne}},\
  }\bibfield  {title} {\enquote {\bibinfo {title} {A quantum-dot
  refrigerator},}\ }\href {\doibase 10.1063/1.110672} {\bibfield  {journal}
  {\bibinfo  {journal} {Appl.\ Phys.\ Lett.}\ }\textbf {\bibinfo {volume}
  {63}},\ \bibinfo {pages} {1815} (\bibinfo {year} {1993})}\BibitemShut
  {NoStop}%
\bibitem [{\citenamefont {Jordan}\ \emph {et~al.}(2013)\citenamefont {Jordan},
  \citenamefont {Sothmann}, \citenamefont
  {S{\ifmmode\acute{a}\else\'{a}\fi}nchez},\ and\ \citenamefont
  {B{\ifmmode\ddot{u}\else\"{u}\fi}ttiker}}]{Jordan2013Feb}%
  \BibitemOpen
  \bibfield  {author} {\bibinfo {author} {\bibfnamefont {Andrew~N.}\
  \bibnamefont {Jordan}}, \bibinfo {author} {\bibfnamefont
  {Bj{\ifmmode\ddot{o}\else\"{o}\fi}rn}\ \bibnamefont {Sothmann}}, \bibinfo
  {author} {\bibfnamefont {Rafael}\ \bibnamefont
  {S{\ifmmode\acute{a}\else\'{a}\fi}nchez}}, \ and\ \bibinfo {author}
  {\bibfnamefont {Markus}\ \bibnamefont
  {B{\ifmmode\ddot{u}\else\"{u}\fi}ttiker}},\ }\bibfield  {title} {\enquote
  {\bibinfo {title} {{Powerful and efficient energy harvester with
  resonant-tunneling quantum dots}},}\ }\href {\doibase
  10.1103/PhysRevB.87.075312} {\bibfield  {journal} {\bibinfo  {journal} {Phys.
  Rev. B}\ }\textbf {\bibinfo {volume} {87}},\ \bibinfo {pages} {075312}
  (\bibinfo {year} {2013})}\BibitemShut {NoStop}%
\bibitem [{\citenamefont {Benenti}\ \emph {et~al.}(2017)\citenamefont
  {Benenti}, \citenamefont {Casati}, \citenamefont {Saito},\ and\ \citenamefont
  {Whitney}}]{Benenti2017Jun}%
  \BibitemOpen
  \bibfield  {author} {\bibinfo {author} {\bibfnamefont {Giuliano}\
  \bibnamefont {Benenti}}, \bibinfo {author} {\bibfnamefont {Giulio}\
  \bibnamefont {Casati}}, \bibinfo {author} {\bibfnamefont {Keiji}\
  \bibnamefont {Saito}}, \ and\ \bibinfo {author} {\bibfnamefont {Robert~S.}\
  \bibnamefont {Whitney}},\ }\bibfield  {title} {\enquote {\bibinfo {title}
  {{Fundamental aspects of steady-state conversion of heat to work at the
  nanoscale}},}\ }\href {\doibase 10.1016/j.physrep.2017.05.008} {\bibfield
  {journal} {\bibinfo  {journal} {Phys. Rep.}\ }\textbf {\bibinfo {volume}
  {694}},\ \bibinfo {pages} {1} (\bibinfo {year} {2017})}\BibitemShut {NoStop}%
\bibitem [{\citenamefont {Cui}\ \emph {et~al.}(2017)\citenamefont {Cui},
  \citenamefont {Miao}, \citenamefont {Jiang}, \citenamefont {Meyhofer},\ and\
  \citenamefont {Reddy}}]{cui2017perspective}%
  \BibitemOpen
  \bibfield  {author} {\bibinfo {author} {\bibfnamefont {Longji}\ \bibnamefont
  {Cui}}, \bibinfo {author} {\bibfnamefont {Ruijiao}\ \bibnamefont {Miao}},
  \bibinfo {author} {\bibfnamefont {Chang}\ \bibnamefont {Jiang}}, \bibinfo
  {author} {\bibfnamefont {Edgar}\ \bibnamefont {Meyhofer}}, \ and\ \bibinfo
  {author} {\bibfnamefont {Pramod}\ \bibnamefont {Reddy}},\ }\bibfield  {title}
  {\enquote {\bibinfo {title} {Perspective: {T}hermal and thermoelectric
  transport in molecular junctions},}\ }\href {\doibase 10.1063/1.4976982}
  {\bibfield  {journal} {\bibinfo  {journal} {J. Chem. Phys.}\ }\textbf
  {\bibinfo {volume} {146}},\ \bibinfo {pages} {092201} (\bibinfo {year}
  {2017})}\BibitemShut {NoStop}%
\bibitem [{\citenamefont {Bedkihal}\ \emph {et~al.}(2025)\citenamefont
  {Bedkihal}, \citenamefont {Behera},\ and\ \citenamefont
  {Bandyopadhyay}}]{bedkihal2025fundamental-review}%
  \BibitemOpen
  \bibfield  {author} {\bibinfo {author} {\bibfnamefont {Salil}\ \bibnamefont
  {Bedkihal}}, \bibinfo {author} {\bibfnamefont {Jayasmita}\ \bibnamefont
  {Behera}}, \ and\ \bibinfo {author} {\bibfnamefont {Malay}\ \bibnamefont
  {Bandyopadhyay}},\ }\bibfield  {title} {\enquote {\bibinfo {title}
  {Fundamental aspects of {Aharonov--Bohm} quantum machines: thermoelectric
  heat engines and diodes},}\ }\href {\doibase 10.1088/1361-648X/adb921}
  {\bibfield  {journal} {\bibinfo  {journal} {J. Phys.: Condens. Matter}\
  }\textbf {\bibinfo {volume} {37}},\ \bibinfo {pages} {163001} (\bibinfo
  {year} {2025})}\BibitemShut {NoStop}%
\bibitem [{\citenamefont {Landauer}(1970)}]{Landauer1970Apr}%
  \BibitemOpen
  \bibfield  {author} {\bibinfo {author} {\bibfnamefont {Rolf}\ \bibnamefont
  {Landauer}},\ }\bibfield  {title} {\enquote {\bibinfo {title} {{Electrical
  resistance of disordered one-dimensional lattices}},}\ }\href {\doibase
  10.1080/14786437008238472} {\bibfield  {journal} {\bibinfo  {journal}
  {Philosophical Magazine: A Journal of Theoretical Experimental and Applied
  Physics}\ }\textbf {\bibinfo {volume} {21}},\ \bibinfo {pages} {863--867}
  (\bibinfo {year} {1970})}\BibitemShut {NoStop}%
\bibitem [{\citenamefont {Engquist}\ and\ \citenamefont
  {Anderson}(1981)}]{Engquist1981Jul}%
  \BibitemOpen
  \bibfield  {author} {\bibinfo {author} {\bibfnamefont {H.-L.}\ \bibnamefont
  {Engquist}}\ and\ \bibinfo {author} {\bibfnamefont {P.~W.}\ \bibnamefont
  {Anderson}},\ }\bibfield  {title} {\enquote {\bibinfo {title} {{Definition
  and measurement of the electrical and thermal resistances}},}\ }\href
  {\doibase 10.1103/PhysRevB.24.1151} {\bibfield  {journal} {\bibinfo
  {journal} {Phys. Rev. B}\ }\textbf {\bibinfo {volume} {24}},\ \bibinfo
  {pages} {1151--1154(R)} (\bibinfo {year} {1981})}\BibitemShut {NoStop}%
\bibitem [{\citenamefont {B{\ifmmode \ddot{u} \else
  \"{u}\fi}ttiker}(1986)}]{Buttiker1986Oct}%
  \BibitemOpen
  \bibfield  {author} {\bibinfo {author} {\bibfnamefont {M.}~\bibnamefont
  {B{\ifmmode \ddot{u} \else \"{u}\fi}ttiker}},\ }\bibfield  {title} {\enquote
  {\bibinfo {title} {{Four-Terminal Phase-Coherent Conductance}},}\ }\href
  {\doibase 10.1103/PhysRevLett.57.1761} {\bibfield  {journal} {\bibinfo
  {journal} {Phys. Rev. Lett.}\ }\textbf {\bibinfo {volume} {57}},\ \bibinfo
  {pages} {1761} (\bibinfo {year} {1986})}\BibitemShut {NoStop}%
\bibitem [{\citenamefont {Pendry}(1983)}]{Pendry1983Jul}%
  \BibitemOpen
  \bibfield  {author} {\bibinfo {author} {\bibfnamefont {J.~B.}\ \bibnamefont
  {Pendry}},\ }\bibfield  {title} {\enquote {\bibinfo {title} {{Quantum limits
  to the flow of information and entropy}},}\ }\href {\doibase
  10.1088/0305-4470/16/10/012} {\bibfield  {journal} {\bibinfo  {journal} {J.
  Phys. A: Math. Gen.}\ }\textbf {\bibinfo {volume} {16}},\ \bibinfo {pages}
  {2161} (\bibinfo {year} {1983})}\BibitemShut {NoStop}%
\bibitem [{\citenamefont {Sivan}\ and\ \citenamefont
  {Imry}(1986)}]{Sivan1986Jan}%
  \BibitemOpen
  \bibfield  {author} {\bibinfo {author} {\bibfnamefont {U.}~\bibnamefont
  {Sivan}}\ and\ \bibinfo {author} {\bibfnamefont {Y.}~\bibnamefont {Imry}},\
  }\bibfield  {title} {\enquote {\bibinfo {title} {{Multichannel Landauer
  formula for thermoelectric transport with application to thermopower near the
  mobility edge}},}\ }\href {\doibase 10.1103/PhysRevB.33.551} {\bibfield
  {journal} {\bibinfo  {journal} {Phys. Rev. B}\ }\textbf {\bibinfo {volume}
  {33}},\ \bibinfo {pages} {551} (\bibinfo {year} {1986})}\BibitemShut
  {NoStop}%
\bibitem [{\citenamefont {Butcher}(1990)}]{Butcher1990}%
  \BibitemOpen
  \bibfield  {author} {\bibinfo {author} {\bibfnamefont {P.~N.}\ \bibnamefont
  {Butcher}},\ }\bibfield  {title} {\enquote {\bibinfo {title} {{Thermal and
  electrical transport formalism for electronic microstructures with many
  terminals}},}\ }\href {\doibase 10.1088/0953-8984/2/22/008} {\bibfield
  {journal} {\bibinfo  {journal} {J. Phys.: Condens. Matter}\ }\textbf
  {\bibinfo {volume} {2}},\ \bibinfo {pages} {4869} (\bibinfo {year}
  {1990})}\BibitemShut {NoStop}%
\bibitem [{\citenamefont {Tesser}\ \emph {et~al.}(2023)\citenamefont {Tesser},
  \citenamefont {Whitney},\ and\ \citenamefont
  {Splettstoesser}}]{tesser2023thermodynamic}%
  \BibitemOpen
  \bibfield  {author} {\bibinfo {author} {\bibfnamefont {Ludovico}\
  \bibnamefont {Tesser}}, \bibinfo {author} {\bibfnamefont {Robert~S}\
  \bibnamefont {Whitney}}, \ and\ \bibinfo {author} {\bibfnamefont {Janine}\
  \bibnamefont {Splettstoesser}},\ }\bibfield  {title} {\enquote {\bibinfo
  {title} {Thermodynamic performance of hot-carrier solar cells: A quantum
  transport model},}\ }\href {\doibase 10.1103/PhysRevApplied.19.044038}
  {\bibfield  {journal} {\bibinfo  {journal} {Phys. Rev. Appl.}\ }\textbf
  {\bibinfo {volume} {19}},\ \bibinfo {pages} {044038} (\bibinfo {year}
  {2023})}\BibitemShut {NoStop}%
\bibitem [{\citenamefont {Bertin-Johannet}\ \emph {et~al.}(2025)\citenamefont
  {Bertin-Johannet}, \citenamefont {Thu\'egaz}, \citenamefont
  {Splettstoesser},\ and\ \citenamefont {Whitney}}]{Bertin2025Improving}%
  \BibitemOpen
  \bibfield  {author} {\bibinfo {author} {\bibfnamefont {Bruno}\ \bibnamefont
  {Bertin-Johannet}}, \bibinfo {author} {\bibfnamefont {Thibaut}\ \bibnamefont
  {Thu\'egaz}}, \bibinfo {author} {\bibfnamefont {Janine}\ \bibnamefont
  {Splettstoesser}}, \ and\ \bibinfo {author} {\bibfnamefont {Robert~S.}\
  \bibnamefont {Whitney}},\ }\bibfield  {title} {\enquote {\bibinfo {title}
  {Improving photovoltaics by adding extra terminals to extract hot
  carriers},}\ }\href {\doibase 10.48550/arXiv.2507.13279} {\bibfield
  {journal} {\bibinfo  {journal} {Preprint}\ } (\bibinfo {year} {2025}),\
  10.48550/arXiv.2507.13279}\BibitemShut {NoStop}%
\bibitem [{\citenamefont
  {B{\ifmmode\ddot{u}\else\"{u}\fi}ttiker}(1990)}]{Buttiker1990Apr}%
  \BibitemOpen
  \bibfield  {author} {\bibinfo {author} {\bibfnamefont {M.}~\bibnamefont
  {B{\ifmmode\ddot{u}\else\"{u}\fi}ttiker}},\ }\bibfield  {title} {\enquote
  {\bibinfo {title} {{Quantized transmission of a saddle-point
  constriction}},}\ }\href {\doibase 10.1103/PhysRevB.41.7906} {\bibfield
  {journal} {\bibinfo  {journal} {Phys. Rev. B}\ }\textbf {\bibinfo {volume}
  {41}},\ \bibinfo {pages} {7906--7909(R)} (\bibinfo {year}
  {1990})}\BibitemShut {NoStop}%
\bibitem [{rev()}]{reviewcomment}%
  \BibitemOpen
  \href@noop {} {}\bibinfo {note} {{See also section 4.4.1 of the review
  article, Ref.~[\onlinecite{Benenti2017Jun}].}}\BibitemShut {Stop}%
\bibitem [{\citenamefont {Kheradsoud}\ \emph {et~al.}(2019)\citenamefont
  {Kheradsoud}, \citenamefont {Dashti}, \citenamefont {Misiorny}, \citenamefont
  {Potts}, \citenamefont {Splettstoesser},\ and\ \citenamefont
  {Samuelsson}}]{Kheradsoud2019Aug}%
  \BibitemOpen
  \bibfield  {author} {\bibinfo {author} {\bibfnamefont {Sara}\ \bibnamefont
  {Kheradsoud}}, \bibinfo {author} {\bibfnamefont {Nastaran}\ \bibnamefont
  {Dashti}}, \bibinfo {author} {\bibfnamefont {Maciej}\ \bibnamefont
  {Misiorny}}, \bibinfo {author} {\bibfnamefont {Patrick~P.}\ \bibnamefont
  {Potts}}, \bibinfo {author} {\bibfnamefont {Janine}\ \bibnamefont
  {Splettstoesser}}, \ and\ \bibinfo {author} {\bibfnamefont {Peter}\
  \bibnamefont {Samuelsson}},\ }\bibfield  {title} {\enquote {\bibinfo {title}
  {Power, efficiency and fluctuations in a quantum point contact as
  steady-state thermoelectric heat engine},}\ }\href {\doibase
  10.3390/e21080777} {\bibfield  {journal} {\bibinfo  {journal} {Entropy}\
  }\textbf {\bibinfo {volume} {21}},\ \bibinfo {pages} {777} (\bibinfo {year}
  {2019})}\BibitemShut {NoStop}%
\bibitem [{\citenamefont {Hajiloo}\ \emph {et~al.}(2020)\citenamefont
  {Hajiloo}, \citenamefont {Alonso}, \citenamefont {Dashti}, \citenamefont
  {Arrachea},\ and\ \citenamefont
  {Splettstoesser}}]{Hajiloo2020Oct-nonlinear-cooling}%
  \BibitemOpen
  \bibfield  {author} {\bibinfo {author} {\bibfnamefont {Fatemeh}\ \bibnamefont
  {Hajiloo}}, \bibinfo {author} {\bibfnamefont
  {Pablo~Terr{\ifmmode\acute{e}\else\'{e}\fi}n}\ \bibnamefont {Alonso}},
  \bibinfo {author} {\bibfnamefont {Nastaran}\ \bibnamefont {Dashti}}, \bibinfo
  {author} {\bibfnamefont {Liliana}\ \bibnamefont {Arrachea}}, \ and\ \bibinfo
  {author} {\bibfnamefont {Janine}\ \bibnamefont {Splettstoesser}},\ }\bibfield
   {title} {\enquote {\bibinfo {title} {{Detailed study of nonlinear cooling
  with two-terminal configurations of topological edge states}},}\ }\href
  {\doibase 10.1103/PhysRevB.102.155434} {\bibfield  {journal} {\bibinfo
  {journal} {Phys. Rev. B}\ }\textbf {\bibinfo {volume} {102}},\ \bibinfo
  {pages} {155434} (\bibinfo {year} {2020})}\BibitemShut {NoStop}%
\bibitem [{\citenamefont {Khomchenko}\ \emph
  {et~al.}(2024{\natexlab{a}})\citenamefont {Khomchenko}, \citenamefont
  {Ouerdane},\ and\ \citenamefont {Benenti}}]{khomchenko2024influence}%
  \BibitemOpen
  \bibfield  {author} {\bibinfo {author} {\bibfnamefont {Ilia}\ \bibnamefont
  {Khomchenko}}, \bibinfo {author} {\bibfnamefont {Henni}\ \bibnamefont
  {Ouerdane}}, \ and\ \bibinfo {author} {\bibfnamefont {Giuliano}\ \bibnamefont
  {Benenti}},\ }\bibfield  {title} {\enquote {\bibinfo {title} {Influence of
  the {A}nderson transition on thermoelectric energy conversion in disordered
  electronic systems},}\ }in\ \href {\doibase 10.1088/1742-6596/2701/1/012018}
  {\emph {\bibinfo {booktitle} {Journal of Physics: Conference Series}}},\
  Vol.\ \bibinfo {volume} {2701}\ (\bibinfo {year} {2024})\ p.\ \bibinfo
  {pages} {012018}\BibitemShut {NoStop}%
\bibitem [{\citenamefont {Khomchenko}\ \emph
  {et~al.}(2024{\natexlab{b}})\citenamefont {Khomchenko}, \citenamefont
  {Ouerdane},\ and\ \citenamefont {Benenti}}]{khomchenko2024corrigendum}%
  \BibitemOpen
  \bibfield  {author} {\bibinfo {author} {\bibfnamefont {Ilia}\ \bibnamefont
  {Khomchenko}}, \bibinfo {author} {\bibfnamefont {Henni}\ \bibnamefont
  {Ouerdane}}, \ and\ \bibinfo {author} {\bibfnamefont {Giuliano}\ \bibnamefont
  {Benenti}},\ }\bibfield  {title} {\enquote {\bibinfo {title} {Corrigendum:
  Influence of the {A}nderson transition on thermoelectric energy conversion in
  disordered electronic systems},}\ }in\ \href {\doibase
  10.1088/1742-6596/2701/1/012149} {\emph {\bibinfo {booktitle} {Journal of
  Physics: Conference Series}}},\ Vol.\ \bibinfo {volume} {2701}\ (\bibinfo
  {year} {2024})\ p.\ \bibinfo {pages} {012149}\BibitemShut {NoStop}%
\bibitem [{\citenamefont {Dubi}\ and\ \citenamefont
  {Di~Ventra}(2011)}]{Dubi2011Mar}%
  \BibitemOpen
  \bibfield  {author} {\bibinfo {author} {\bibfnamefont {Yonatan}\ \bibnamefont
  {Dubi}}\ and\ \bibinfo {author} {\bibfnamefont {Massimiliano}\ \bibnamefont
  {Di~Ventra}},\ }\bibfield  {title} {\enquote {\bibinfo {title} {{Colloquium:
  Heat flow and thermoelectricity in atomic and molecular junctions}},}\ }\href
  {\doibase 10.1103/RevModPhys.83.131} {\bibfield  {journal} {\bibinfo
  {journal} {Rev. Mod. Phys.}\ }\textbf {\bibinfo {volume} {83}},\ \bibinfo
  {pages} {131--155} (\bibinfo {year} {2011})}\BibitemShut {NoStop}%
\bibitem [{\citenamefont {Sowa}\ \emph {et~al.}(2019)\citenamefont {Sowa},
  \citenamefont {Mol},\ and\ \citenamefont {Gauger}}]{sowa2019marcus}%
  \BibitemOpen
  \bibfield  {author} {\bibinfo {author} {\bibfnamefont {Jakub~K}\ \bibnamefont
  {Sowa}}, \bibinfo {author} {\bibfnamefont {Jan~A}\ \bibnamefont {Mol}}, \
  and\ \bibinfo {author} {\bibfnamefont {Erik~M}\ \bibnamefont {Gauger}},\
  }\bibfield  {title} {\enquote {\bibinfo {title} {Marcus theory of
  thermoelectricity in molecular junctions},}\ }\href {\doibase
  10.1021/acs.jpcc.8b12163} {\bibfield  {journal} {\bibinfo  {journal} {J.
  Phys.\ Chem.\ C}\ }\textbf {\bibinfo {volume} {123}},\ \bibinfo {pages}
  {4103} (\bibinfo {year} {2019})}\BibitemShut {NoStop}%
\bibitem [{\citenamefont {Kirchberg}\ and\ \citenamefont
  {Nitzan}(2022)}]{kirchberg2022energy}%
  \BibitemOpen
  \bibfield  {author} {\bibinfo {author} {\bibfnamefont {Henning}\ \bibnamefont
  {Kirchberg}}\ and\ \bibinfo {author} {\bibfnamefont {Abraham}\ \bibnamefont
  {Nitzan}},\ }\bibfield  {title} {\enquote {\bibinfo {title} {Energy transfer
  and thermoelectricity in molecular junctions in non-equilibrated solvents},}\
  }\href {\doibase 10.1063/5.0086319} {\bibfield  {journal} {\bibinfo
  {journal} {J. Chem.\ Phys.}\ }\textbf {\bibinfo {volume} {156}} (\bibinfo
  {year} {2022}),\ 10.1063/5.0086319}\BibitemShut {NoStop}%
\bibitem [{\citenamefont {Sothmann}\ \emph {et~al.}(2013)\citenamefont
  {Sothmann}, \citenamefont {S{\ifmmode\acute{a}\else\'{a}\fi}nchez},
  \citenamefont {Jordan},\ and\ \citenamefont
  {B{\ifmmode\ddot{u}\else\"{u}\fi}ttiker}}]{Sothmann2013Sep}%
  \BibitemOpen
  \bibfield  {author} {\bibinfo {author} {\bibfnamefont
  {Bj{\ifmmode\ddot{o}\else\"{o}\fi}rn}\ \bibnamefont {Sothmann}}, \bibinfo
  {author} {\bibfnamefont {Rafael}\ \bibnamefont
  {S{\ifmmode\acute{a}\else\'{a}\fi}nchez}}, \bibinfo {author} {\bibfnamefont
  {Andrew~N.}\ \bibnamefont {Jordan}}, \ and\ \bibinfo {author} {\bibfnamefont
  {Markus}\ \bibnamefont {B{\ifmmode\ddot{u}\else\"{u}\fi}ttiker}},\ }\bibfield
   {title} {\enquote {\bibinfo {title} {{Powerful energy harvester based on
  resonant-tunneling quantum wells}},}\ }\href {\doibase
  10.1088/1367-2630/15/9/095021} {\bibfield  {journal} {\bibinfo  {journal}
  {New J. Phys.}\ }\textbf {\bibinfo {volume} {15}},\ \bibinfo {pages} {095021}
  (\bibinfo {year} {2013})}\BibitemShut {NoStop}%
\bibitem [{Note2()}]{Note2}%
  \BibitemOpen
  \bibinfo {note} {Maximum efficiencies always grow as $T_{\protect \rm
  L}/T_{\protect \rm R}$ increases, so it is assumed that $T_{\protect \rm
  L}/T_{\protect \rm R}$ takes the largest value that is achievable in the
  context.}\BibitemShut {Stop}%
\bibitem [{Note3()}]{Note3}%
  \BibitemOpen
  \bibinfo {note} {There is a direct analogy here with a household
  refrigerator. There the heat leaks are often dominated by heat flow through
  the insulating material around the cold compartment. This heat flow is
  unrelated to the properties of the refrigeration circuit.}\BibitemShut
  {Stop}%
\bibitem [{Note4()}]{Note4}%
  \BibitemOpen
  \bibinfo {note} {Scattering theory relies on this assumption; this
  assumption's legitimacy comes from this theory often describing experimental
  observations. However, there are also systems where scattering theory is
  inapplicable because of strong electron-phonon interactions. We mention such
  issues in our conclusions.}\BibitemShut {Stop}%
\bibitem [{Note5()}]{Note5}%
  \BibitemOpen
  \bibinfo {note} {For example, see section 6.1 of the review article,
  Ref.~[\protect \rev@citealpnum {Benenti2017Jun}].}\BibitemShut {Stop}%
\bibitem [{\citenamefont {Brandner}\ and\ \citenamefont
  {Seifert}(2015)}]{Brandner2015Jan}%
  \BibitemOpen
  \bibfield  {author} {\bibinfo {author} {\bibfnamefont {Kay}\ \bibnamefont
  {Brandner}}\ and\ \bibinfo {author} {\bibfnamefont {Udo}\ \bibnamefont
  {Seifert}},\ }\bibfield  {title} {\enquote {\bibinfo {title} {{Bound on
  thermoelectric power in a magnetic field within linear response}},}\ }\href
  {\doibase 10.1103/PhysRevE.91.012121} {\bibfield  {journal} {\bibinfo
  {journal} {Phys. Rev. E}\ }\textbf {\bibinfo {volume} {91}},\ \bibinfo
  {pages} {012121} (\bibinfo {year} {2015})}\BibitemShut {NoStop}%
\bibitem [{\citenamefont {Luo}\ \emph {et~al.}(2018)\citenamefont {Luo},
  \citenamefont {Benenti}, \citenamefont {Casati},\ and\ \citenamefont
  {Wang}}]{Luo2018Aug}%
  \BibitemOpen
  \bibfield  {author} {\bibinfo {author} {\bibfnamefont {Rongxiang}\
  \bibnamefont {Luo}}, \bibinfo {author} {\bibfnamefont {Giuliano}\
  \bibnamefont {Benenti}}, \bibinfo {author} {\bibfnamefont {Giulio}\
  \bibnamefont {Casati}}, \ and\ \bibinfo {author} {\bibfnamefont {Jiao}\
  \bibnamefont {Wang}},\ }\bibfield  {title} {\enquote {\bibinfo {title}
  {Thermodynamic bound on heat-to-power conversion},}\ }\href {\doibase
  10.1103/PhysRevLett.121.080602} {\bibfield  {journal} {\bibinfo  {journal}
  {Phys. Rev. Lett.}\ }\textbf {\bibinfo {volume} {121}},\ \bibinfo {pages}
  {080602} (\bibinfo {year} {2018})}\BibitemShut {NoStop}%
\bibitem [{\citenamefont {Chrirou}\ \emph {et~al.}(2026)\citenamefont
  {Chrirou}, \citenamefont {El~Allati},\ and\ \citenamefont
  {Whitney}}]{zenodo-codes}%
  \BibitemOpen
  \bibfield  {author} {\bibinfo {author} {\bibfnamefont {Chaimae}\ \bibnamefont
  {Chrirou}}, \bibinfo {author} {\bibfnamefont {Abderrahim}\ \bibnamefont
  {El~Allati}}, \ and\ \bibinfo {author} {\bibfnamefont {Robert~S.}\
  \bibnamefont {Whitney}},\ }\href@noop {} {} (\bibinfo {year} {2026}),\
  \bibinfo {note} {\textit{``Python codes and data for arXiv:2507.14977''}
  \href{https://doi.org/10.5281/zenodo.18983476}{doi.org/10.5281/zenodo.18983476}}\BibitemShut
  {NoStop}%
\end{thebibliography}%
\end{document}